\pdfoutput=1
\documentclass[acmtog]{acmart}
\acmSubmissionID{776}
\AtBeginDocument{%
  \providecommand\BibTeX{{%
    \normalfont B\kern-0.5em{\scshape i\kern-0.25em b}\kern-0.8em\TeX}}}

\newcommand{\modelfull}{Neural Flow Maps\xspace}
\newcommand{\model}{NFM\xspace}
\newcommand{\inrfull}{Spatially Sparse Neural Fields\xspace}
\newcommand{\inr}{SSNF\xspace}

\setcopyright{acmlicensed}
\acmJournal{TOG}
\acmYear{2023} \acmVolume{42} \acmNumber{6} \acmArticle{244} \acmMonth{12} \acmPrice{15.00}\acmDOI{10.1145/3618392}

\definecolor{MyRed}{rgb}{0.65,0.07,0.09}

\definecolor{MyGreen}{rgb}{0.18,0.55,0.09}

\usepackage{wrapfig}
\usepackage{diagbox}
\usepackage{verbatim}
\usepackage{amssymb}
\usepackage{latexsym}
\usepackage{lineno}
\usepackage{xspace}
\usepackage{color}
\graphicspath{{figures/}}
\usepackage{amsmath,bm}
\usepackage{psfrag}
\usepackage{mathtools}

\theoremstyle{definition}

\usepackage{multirow, tabularx}
\newcolumntype{Y}{>{\centering\arraybackslash}X}
\usepackage{capt-of}%
\usepackage{array, boldline, makecell, booktabs}

\newcommand{\Mod}[1]{\ (\mathrm{mod}\ #1)}

\usepackage{array}

\usepackage{pifont}%

\usepackage{psfrag}
\usepackage{makecell}
\usepackage{algorithm}
\usepackage{threeparttable,booktabs}
\usepackage{graphicx}
\usepackage{subcaption}
\usepackage{flushend}

\usepackage{algpseudocode}

\newtheorem*{remark}{Remark}

\usepackage{xcolor}
\usepackage{graphicx}

\newcommand{\rev}[1]{{#1}}

\newif\ifverbose
\verbosetrue

\ifverbose

\newcommand{\vb}[1]{\textcolor{red}{#1}}
\else

\newcommand{\vb}[1]{}
\fi

\usepackage{url}
\usepackage{xcolor}
\definecolor{newcolor}{rgb}{.8,.349,.1}

\usepackage{float}
\floatstyle{plaintop}
\restylefloat{table}

\begin{document}

\author{Yitong Deng}
\affiliation{
\institution{Dartmouth College}
\state{USA}
}
\email{yitong.deng.gr@dartmouth.edu}

\author{Hong-Xing Yu}
\affiliation{
\institution{Stanford University}
\state{USA}
}
\email{koven@cs.stanford.edu}

\author{Diyang Zhang}
\affiliation{
\institution{Dartmouth College}
\state{USA}
}
\email{diyang.zhang.gr@dartmouth.edu}

\author{Jiajun Wu}
\affiliation{
\institution{Stanford University}
\state{USA}
}
\email{jiajunwu@cs.stanford.edu}

\author{Bo Zhu}
\affiliation{
\institution{Georgia Institute of Technology, Dartmouth College}
\state{USA}
}
\email{bo.zhu@gatech.edu}

\title{Fluid Simulation on Neural Flow Maps}

\newcommand{\revise}[1]{{#1}}
\newcommand{\revisea}[1]{{#1}}

\begin{abstract}
We introduce \modelfull, a novel simulation method bridging the emerging paradigm of implicit neural representations with fluid simulation based on the theory of flow maps, to achieve state-of-the-art simulation of inviscid fluid phenomena. 
We devise a novel hybrid neural field representation, \inrfull (\inr), which fuses small neural networks with a pyramid of overlapping, multi-resolution, and spatially sparse grids, to compactly represent long-term spatiotemporal velocity fields at high accuracy. 
With this neural velocity buffer in hand, we compute long-term, bidirectional flow maps and their Jacobians in a mechanistically symmetric manner, to facilitate drastic accuracy improvement over existing solutions. These long-range, bidirectional flow maps enable high advection accuracy with low dissipation, which in turn facilitates high-fidelity incompressible flow simulations that manifest intricate vortical structures. 
We demonstrate the efficacy of our neural fluid simulation in a variety of challenging simulation scenarios, including leapfrogging vortices, colliding vortices, vortex reconnections, as well as vortex generation from moving obstacles and density differences. Our examples show increased performance over existing methods in terms of energy conservation, visual complexity, adherence to experimental observations, and preservation of detailed vortical structures.
\end{abstract}



\begin{teaserfigure}
 \centering
 \includegraphics[width=.99\textwidth]{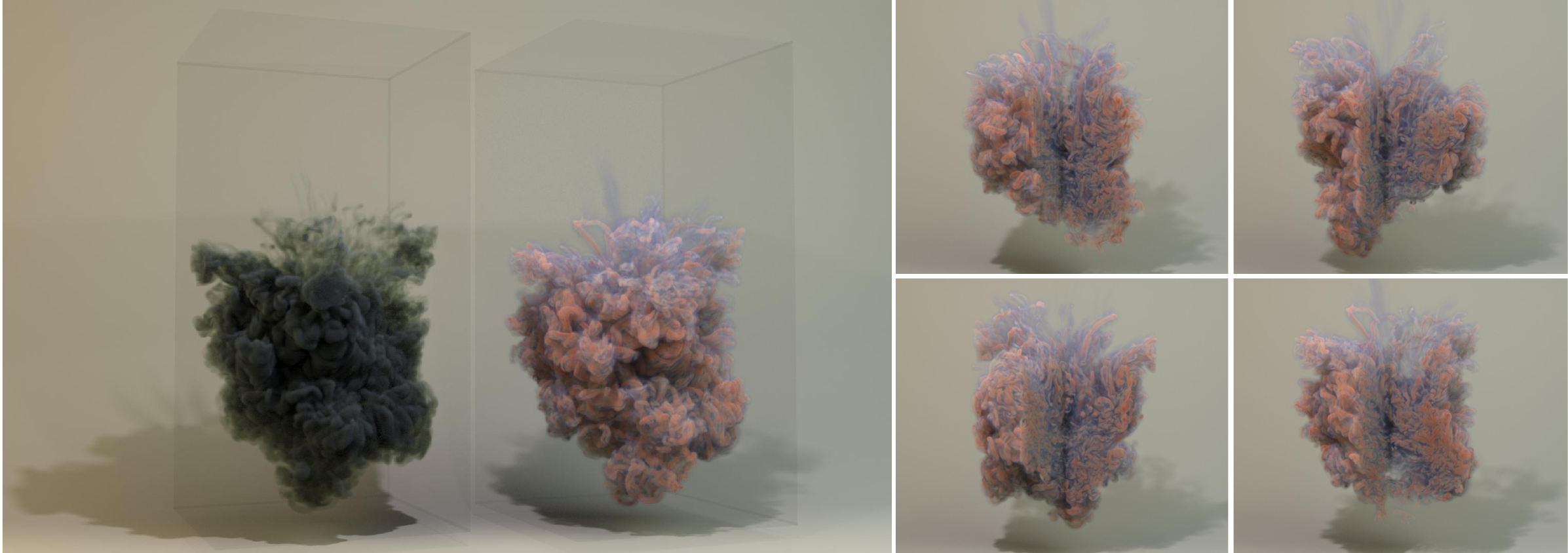}
 \caption{Inkdrop simulated by \modelfull, a novel, high-fidelity simulation system built upon implicit neural representations. The left panel juxtaposes the ink (left) and the underlying bundle of vortex filaments (right). The right panel shows the vorticity \rev{(magnitude)} field from four different angles, with the camera-facing sections culled to reveal the sophisticated internal structures.}
 \label{fig:teaser}
\end{teaserfigure}

\maketitle
\section{Introduction}
Implicit neural representations (INR) have emerged as a remarkable new class of data primitives in visual computing, as an attractive alternative to traditional, explicit representations such as grids, meshes, and particles, demonstrating its distinct advantages in representing various categories of continuous fields in appearance modeling, geometry processing, and 3D reconstruction, just to name a few. Instead of representing a field by storing a large number of samples at discrete locations, an INR employs a continuous query function in the form of neural networks that evaluates the field at arbitrary input coordinates in the domain. Such an alternative representation has proven to be uniquely advantageous in a variety of scenarios for being spatially adaptive, differentiable, agnostic to topology and dimensionality, and memory efficient; and it has empowered many state-of-the-art advances in 3D computer vision (e.g., Instant~NGP~\citep{muller2022instant}). 
One fundamental reason for this is that INRs are highly efficient and adaptive in memory, as it scales up with the \textit{complexity} of the signals instead of the \textit{resolution}~\citep{sitzmann2019scene}, hence allowing the high-fidelity representation of high-dimensional signals such as 5D Neural Radiance Fields~\citep{mildenhall2020nerf}.
In spite of its typical association with radiance fields in rendering and reconstruction tasks, INRs are developing into a fundamentally general field representation paradigm whose virtues promise to be leveraged in a plethora of other disciplines.

Physical simulation, on the other hand, is a discipline in which fields are first-class citizens. When devising a simulation algorithm, a foundational first step is to design an appropriate representation for the dynamical fields: \textit{e.g.}, velocity, pressure, density, and temperature, on which the numerical solving of PDEs can take place. While the options are traditionally different flavors of meshes and particles, the rapid development of INR elicits the curiosity of whether the adoption of such a novel implicit paradigm can fundamentally unlock new horizons of what physical simulation algorithms can achieve.
Unfortunately, the \textit{drop-in} incorporation of INRs in the simulation pipeline has not yet been lucrative. 
A major reason for this is that the principal advantage of INRs, \textit{i.e.}, their spatially adaptive, dimensionality-agnostic, high memory efficiency, 
does not exhibit clear advantages in a typical physical simulation pipeline, which operates on dense fields with only spatial dimensions.
Therefore, dynamical simulation using INRs has so far remained a ``workable concept'' that nevertheless shows no concrete advantage over existing paradigms.
From our perspective, this conundrum is fundamentally caused by the current \textit{mismatch} between the simulation algorithm's demands and the INR's capacities. As with grids, meshes, and particles, an INR's advantages and drawbacks need to be carefully aligned with the simulation scheme to fulfill its full potential. 

In this work, we consolidate INRs' potency in encoding spatiotemporal signals at a small memory footprint with flow map-based fluid advection, an accurate and non-dissipative advection scheme that has been hindered by its intractable memory requirement~\citep{sato2018spatially}. To fulfill the full potential of flow map advection at a viable memory cost, we propose \modelfull (\model), a novel simulation method that realizes accurate, long-range flow maps with neural representations. The core component of \model is the \inrfull (\inr), a hybrid INR employing a pyramid of multi-resolution, spatially sparse feature grids to maintain long-range buffers of spatiotemporal velocity fields. We show that for this purpose, our \inr offers improved accuracy, training speed, and memory efficiency compared to state-of-the-art representations (\textit{e.g.}, Instant~NGP~\cite{muller2022instant}, K-Planes~\cite{fridovich2023k}, and SIREN~\cite{sitzmann2020implicit}), reducing the fitting error by over 70\%. Leveraging this effective neural structure, we compute high-quality bidirectional flow maps by marching the \inr forward and backward in time, and consolidate them with the impulse-based fluid simulation method to drastically reduce simulation errors by over 90\% with respect to analytical solutions. 
\rev{In terms of practicality, our neural method is still significantly slower than traditional methods like Stable Fluids~\cite{stam1999stable}, typically by an order of magnitude, rendering it unsuited for real-time applications, albeit realistic for quality-focused applications, a realm in which our method brings substantial advantages.} These advantages are experimentally validated in a variety of challenging scenarios including leapfrogging vortices, vortex reconnections, and vortex generation from obstacles and density difference, in which our method consistently achieves the state-of-the-art level of energy conservation, visual complexity, and recreation of real-world phenomena. 
\rev{Since the cost of our cutting-edge results hinges on the efficiency of neural techniques, our method effectively reformulates a long-standing simulation problem into a machine-learning one, thereby opening up brand new avenues for the former that promise to leverage the latter's revolutionary advancements of late.}

We summarize our main contributions as follows:
\begin{enumerate}
    \item We present \modelfull (\model), the first INR-based dynamic simulation method to achieve state-of-the-art performance in terms of both visual quality and physical fidelity.
    \item We propose \inrfull (\inr), a novel hybrid INR that is fast to train, compact, and collision-free. Our \inr achieves higher fitting accuracy than the state-of-the-art methods at a lower computational cost.
    \item We introduce a novel scheme for learning neural buffers of long-term spatiotemporal signals, from \textit{streams} of input frames with variable durations.
    \item We introduce a novel flow map advection scheme based on the forward and backward marching of velocity buffers to represent accurate and consistent bidirectional mappings and their Jacobians to enable high-quality fluid simulations.
\end{enumerate}

\section{Related works}

\subsection{Machine Learning for Fluids}
Machine learning has been successfully applied to the study of fluids in a number of directions, as we will briefly survey.
Readers can refer to comprehensive surveys \cite{brunton2020machine,sharma2023review,chen2021data,majchrzak2023survey} for in-depth discussions.
\paragraph{Synthesis}
Machine learning can synthesize visually appealing details from coarse simulations. For instance, \citet{chu2017data} adopt convolutional neural networks to enhance coarse smoke simulations with pre-computed patches; \citet{kim2019deep} synthesize divergence-free fluid motion in a reduced space; \citet{xie2018tempogan} and \citet{werhahn2019multi} use generative adversarial networks to perform fluid super-resolution, an approach adapted by \citet{chu2021learning} to enable user control. \rev{\citet{roy2021neural} upsample particle-based simulations by learning a deformation field using an architecture adapted from FlowNet3D \cite{liu2019flownet3d}.} 
Physics-inspired syntheses of \textit{stylized} fluid details have also been accomplished in the Eulerian~\cite{kim2019transport} and Lagrangian~\cite{kim2020lagrangian} settings.

\paragraph{Inference}
Early works~\cite{gregson2012stochastic, hasinoff2007photo, ihrke2004image} infer density volumes from videos, while more recent ones further infer velocity using physical priors~\citep{gregson2014capture,okabe2015fluid, eckert2018coupled, eckert2019scalarflow,zang2020tomofluid, franz2021global}. Physics-informed neural networks use soft constraints to infer velocity from density~\citep{raissi2019physics,raissi2020hidden}, to which \citet{chu2022physics} combine differentiable rendering to \textit{end-to-end} infer flow from videos. 
\citet{deng2023learning} introduce neural vortex modeling to infer fluid motion and predict future evolution.
\rev{\citet{jakob} and \citet{sahoo2022neural} use neural techniques to infer high-quality flow maps from sparse samples. While also bridging neural and flow-map concepts, their methods are for post-processing \textit{existing simulations}, while we tackle a fundamentally different problem with a powerful neural \textit{simulator}.} 

\paragraph{Inverse and Control Problems}
Learned fluid models can also benefit computational design and control tasks. \citet{wandel2020learning} control the vortex shedding frequency based on a learned physics-informed network, which is extended by \citet{nava2022fast} to control a 2D soft swimmer. \citet{li2018learning} learn particle-based simulators with gradient-based trajectory optimization for fluid manipulation, while \citet{li20213d} learn 3D fluid scenes based on 2D observations to enable gradient-free model-predictive control.

\paragraph{Acceleration}
Data-driven simulators can emulate traditional simulators at a reduced time cost, by replacing expensive iterative solvers with more efficient operations like convolution \cite{tompson2017accelerating} \rev{and message passing \cite{li2022graph}}, capturing fine-grain details on coarse \rev{resolutions} \cite{stachenfeld2021learned, kochkov2021machine}, learning reduced-order latent spaces \cite{kim2019deep, wiewel2019latent,wiewel2020latent}, or lifting the timestep restriction by learning long-term state transitions across multiple simulation timesteps with graph neural networks \cite{pfaff2020learning, sanchez2020learning}, spatial convolution \cite{ummenhofer2020lagrangian}, or regression forests \cite{ladicky2015data}. These methods enable numerical accuracy to be traded for computational savings.

\paragraph{Physical Accuracy}
Machine learning can also enhance the accuracy of traditional methods, using data to supplement the insufficiently modeled dynamics in, for instance, the Reynolds Averaged Navier–Stokes (RANS) method \cite{majchrzak2023survey}, correcting the errors caused by the closure coefficients and Reynolds stress anisotropy with extended kernel regression \cite{tracey2013application}, random forest \cite{wang2017physics}, field inversion and machine learning \cite{duraisamy2015new}, and neural networks \cite{singh2017machine}.

\subsection{Implicit Neural Representation}
INRs have been used widely in visual computing applications including geometry processing~\citep{park2019deepsdf,mescheder2019occupancy}, image-based rendering~\citep{sitzmann2019scene}, and inverse rendering~\citep{yu2023learning,zhang2021physg}, for being spatially adaptive, resolution-agnostic, and memory efficient. One seminal work is Neural Radiance Fields (NeRF)~\citep{mildenhall2020nerf} which represents a 5D radiance field with a neural network that only occupies a few tens of megabytes. 
However, pure neural representations suffer from their considerable time cost, and follow-up works have focused on hybrid INRs featuring classical data structures like dense~\citep{sun2022direct} and sparse voxel grids\rev{~\citep{liu2020neural,chabra2020deep, takikawa2021neural,martel2021acorn,yu2021plenoctrees, kim2022neuralvdb}}. Recently, plane-based data structures~\citep{fridovich2023k,chen2022tensorf} have also been leveraged to good effects. Most relevant to our work is Instant~NGP~\citep{muller2022instant}, which uses multi-resolution hashing for fast training, but lacks both the explicit treatment of hash collisions and temporal modeling. 
Recent works have also used INRs to represent dynamic scenes, by learning deformation~\citep{park2021hypernerf,pumarola2021d,park2021nerfies} and scene flow fields~\citep{li2021neural,du2021neural,xian2021space,li2022neural}. 

\newcolumntype{z}{X}
\newcolumntype{s}{>{\hsize=.25\hsize}X}
\setlength{\abovecaptionskip}{5pt}
\begin{table}
\centering\small
\begin{tabularx}{0.47\textwidth}{ssz}
\hlineB{3}
Notation & Type & Definition\\
\hlineB{2.5}
\hspace{12pt}$X$ & vector & material point position at initial state\\
\hlineB{1}
\hspace{12pt}$x$ & vector & material point position at current state\\
\hlineB{1}
\hspace{12pt}$\phi$ & vector & forward map\\
\hlineB{1}
\hspace{12pt}$\psi$ & vector & backward map\\
\hlineB{1}
\hspace{12pt}$\mathcal{F}$ & matrix & forward map gradients\\
\hlineB{1}
\hspace{12pt}$\mathcal{T}$ & matrix & backward map gradients\\
\hlineB{1}
\hspace{12pt}$\bm{u}$ & vector & velocity\\
\hlineB{1}
\hspace{12pt}$\bm{m}$ & vector & impulse\\
\hlineB{1}
\hspace{12pt}$\mathcal{N}$ & \inr & neural buffer\\
\hlineB{1}
\hspace{12pt}$S$ & scalar & sizing field\\
\hlineB{1}
\hspace{12pt}$n$ & scalar & reinitialization steps\\
\hlineB{3}
\end{tabularx}
\caption{Symbols and notations used in this paper.}
\vspace{-6pt}
\label{tab: notation_table}
\end{table}

\subsection{Fluid Simulation}
\rev{
The pioneering works by \citet{Foster1997ModelingTM} and \citet{stam1999stable} lay the groundwork for fluid simulation in computer graphics with the employment of uniform Marker-and-Cell (MAC) grids \cite{harlow1965numerical}, Chorin's projection method \cite{chorin1968numerical}, and the semi-Lagrangian advection scheme \cite{Sawyer1963ASM,Robert1981ASN}. Since then, researchers have developed upon this paradigm on multiple fronts to improve on the efficiency and accuracy.
}

\paragraph{Hierarchical Representation}
\rev{
Compared to uniform discretization, hierarchical or multi-resolution discretizations can capture the intricate details more efficiently by biasing the computational resources towards regions of interest, which has motivated researchers to design computational systems using nested layers of uniform grids \cite{setaluri2014spgrid, Minion1994TwoMF, Martin1996SolvingPE, Martin2007ACA, Johansen1998ACG}, Chimera grids \cite{Henshaw1994AFA, English2013AnAD}, and non-uniform grids like quadtrees and octrees \cite{Popinet2003GerrisAT,Losasso2004SimulatingWA,Losasso2006SpatiallyAT,Batty2017ACF,ando2020practical}. In addition, multi-resolution modeling has also been achieved with wavelet \cite{Deriaz2006DivergencefreeAC, Kevlahan2005AnAW, Schneider2010WaveletMI} and model-reduced methods \cite{mercier2020local}. 
}

\paragraph{Numerical Dissipation Reduction} \rev{The Stable Fluids method \cite{stam1999stable} yields a significant amount of numerical viscosity causing the results to appear blurred and damped. Researchers have addressed this artifact with error correction schemes \cite{kim2006advections, selle2008unconditionally}, higher-order interpolation \cite{losasso2006spatially, Nave2009AGL}, improved backtracking schemes \cite{Jameson1981NumericalSO, cho2018dual}, particle-based advection \cite{zhu2005animating, jiang2015affine, Fu2017APP}, energy-preserving integration, \cite{Mullen2009EnergypreservingIF}, \textit{a posteriori} vorticity correction \cite{fedkiw2001visual, Zhang2015RestoringTM}, reflection \cite{zehnder2018advection, narain2019second, Strang1968OnTC} and Lie advection \cite{nabizadeh2022covector}. Flow map methods offer another alternative option, as elaborated below.}

\paragraph{Flow Map Methods}
The method of characteristic mapping (MCM) proposed by \citet{wiggert1976numerical} pioneers in solving the fluid equations on flow maps, a technique introduced to the graphics community by 
\citet{tessendorf2011characteristic}. Tracking long-term flow maps instead of fluid quantities, MCM drastically reduces the number of interpolations and hence the resulting numerical diffusion.
We refer to \rev{the work by \citet{tessendorf2015advection}} for more in-depth analyses and derivations on MCM. Earlier methods \cite{sato2018spatially, Sato2017ALS, tessendorf2011characteristic, hachisuka2005combined} compute flow maps by tracing virtual particles which leads to a demanding time and memory cost. \citet{qu2019efficient} trade off accuracy for efficiency by advecting the backward map in a semi-Lagrangian-like manner; and bring in the forward map to realize the BFECC error compensation \cite{kim2006advections}. 
\citet{nabizadeh2022covector} effectively combine bidirectional flow maps with impulse-based fluid dynamics \cite{Oseledets1989COMMUNICATIONSOT, Buttke1992LagrangianNM,feng2022impulse}, using the mappings \textit{along with their Jacobians} to perform circulation-preserving advection. 

\section{Physical Model}
\subsection{Mathematical Foundation}

\paragraph{Flow Map Preliminaries}

We start by defining a spatiotemporal velocity field $\bm u(\bm q,\tau)$ in the fluid domain $\Omega$ to specify the velocity vector at location $\bm q$ and time $\tau$. For a material point $\bm X \in \Omega$, the corresponding forward flow map $\phi:\Omega_0\rightarrow\Omega_t$ is defined as:
\begin{equation}
    \label{eq:phi_def}
    \begin{dcases}
    \frac{\partial \phi(\bm X,\tau)}{\partial \tau}=\bm u(\phi(\bm X,\tau),\tau),\\
    \phi(\bm X,0) = \bm X, \\
    \phi(\bm X,t) = \bm x,
    \end{dcases}
\end{equation}
in which $\bm X$ represents the material point's position at time 0, and $\bm x$ specifies the position of the same material point at time $t$. The forward mapping from $\bm X$ to $\bm x$ is determined by the trajectory of $\bm X$, which is calculated by integrating $\bm u$ within the time interval $[0, t]$.

Similarly, we define the backward flow map  $\psi:\Omega_t\rightarrow\Omega_{0}$ as:
\begin{equation}
    \begin{dcases}
    \frac{\partial \psi(\bm x,\tau)}{\partial \tau}=\bm u(\psi(\bm x,\tau),\tau),\\
    \psi(\bm x,t) = \bm x, \\
    \psi(\bm x,0) = \bm X.
    \end{dcases}
    \label{eq:flow_map}
\end{equation}
Consequently, integrating $\bm u$ reversely from time $t$ to time $0$ on the temporal axis, starting from $\bm x$ in the spatial domain, yields the backward mapping from $\bm x$ to $\bm X$. Without causing confusion, in the following discussion, we will abbreviate $\phi(\bm X,t)$ as $\phi$ and $\psi(\bm x,0)$ as $\psi$. We can also use $\phi$ and $\psi$ interchangeably with $\bm x$ and $\bm X$.

\begin{figure}
 \centering
 \includegraphics[width=.40\textwidth]{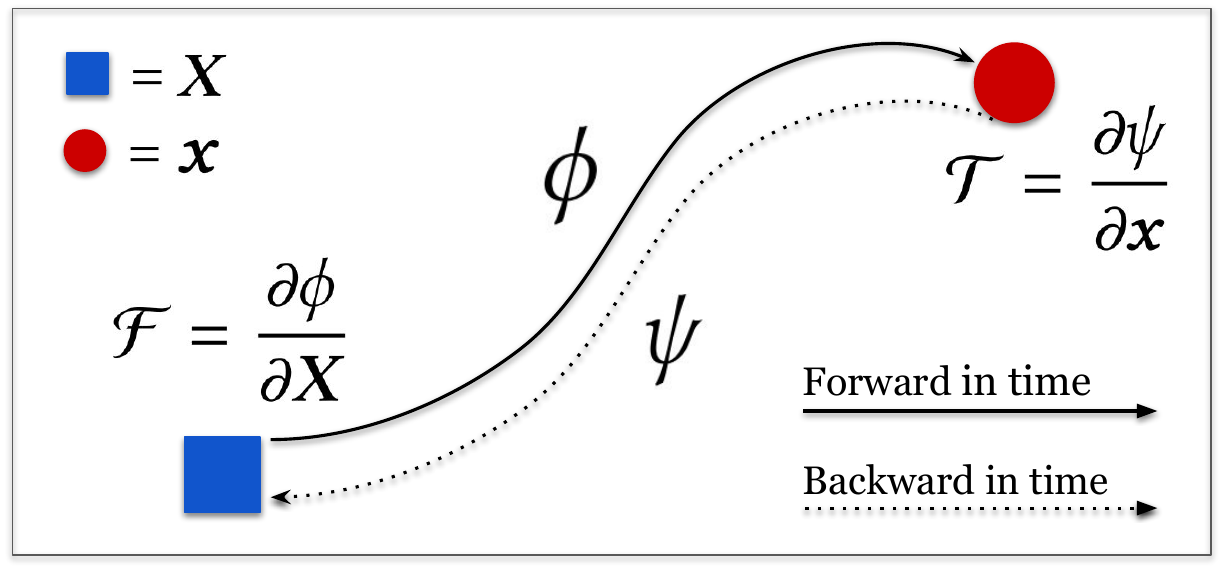}
 \caption{The forward map $\phi$ sends $\bm X$ (blue square) to $\bm x$ (red circle) by traveling forward in time; the backward map $\psi$ sends $\bm x$ to $\bm X$ by traveling backward in time. The Jacobians $\mathcal{F}$ and $\mathcal{T}$ are calculated at $\bm X$ and $\bm x$ respectively.}
 \label{fig:flowmap_101}
 \vspace{-1pt}
\end{figure}

As seen in Figure~\ref{fig:flowmap_101}, intuitively, the mapping $\phi$ sends a material point to its current position at time $t$ given its initial position; the mapping $\psi$ backtracks a material from its current position at time $t$ to its initial position. 

We denote the spatial gradients (Jacobians) of $\phi$ and $\psi$ with $\mathcal{F}$ and $\mathcal{T}$ as:
\begin{equation}
    \begin{aligned}
        \mathcal{F} = \frac{\partial \phi}{\partial \bm X}, \\
        \mathcal{T} = \frac{\partial \psi}{\partial \bm x}.
    \end{aligned}
    \label{eq:F_def}
\end{equation}

Symbolically, $\mathcal{F}$ and $\mathcal{T}$ represent the deformation between the initial and current reference frames. In continuum mechanics, $\mathcal{F}$ is typically known as the deformation gradient. The equations of the temporal evolution of $\mathcal{F}$ and $\mathcal{T}$ are given by: 
\begin{equation}
    \begin{aligned}
        \frac{D \mathcal{F}}{D t}&=\nabla\bm{u}\,\mathcal{F}, \label{eq:F_evolution}\\
        \frac{D \mathcal{T}}{D t}&=-\mathcal{T}\,\nabla\bm{u}.
    \end{aligned}
\end{equation}
Both equations are written in matrix form. We refer readers to \rev{the book by \citet{firstcourse2008}} for a more detailed derivation.

\setlength{\abovecaptionskip}{12pt}

\paragraph{Flow Maps for Fluid Simulation}
Since the four flow map quantities: $\phi$, $\psi$, $\mathcal{F}$ and $\mathcal{T}$ can fully prescribe material transport, they can serve as essential ingredients for accurate fluid simulations. 
To illustrate, we consider the impulse form of the Euler equations for inviscid flow \cite{feng2022impulse,nabizadeh2022covector,cortez1995impulse}:
\begin{equation}
\begin{dcases}
 \frac{D \bm m }{D t} =-\left(\nabla \bm u\right)^{T}\,\bm m,\\
\nabla^2 \varphi = \nabla \cdot \bm m,\\
\bm u = \bm m - \nabla \varphi,
\end{dcases}
\label{eq:m_NS}
\end{equation}
with $\bm m$ and $\bm u$ being fluid impulse and velocity, and $\varphi$ an intermediate variable used only for projecting $\bm m$ to the divergence-free $\bm u$. The impulse $\bm m$ is considered to contain more information than $\bm u$, as $\bm u$ can be reconstructed from $\bm m$ by solving the Poisson equation.

A flow map-based solver for Equation~\ref{eq:m_NS} is introduced to the graphics community by \citet{nabizadeh2022covector}, which solves the equation by first computing the backward flow map $\psi$ and its Jacobian $\mathcal{T}$, and then reconstructing the impulse $\bm m$ by evaluating:
\begin{equation}
    \bm m(\bm x,t)=\mathcal{T}^{T}\,\bm{m}(\psi(\bm x),0).
    \label{eq:imp_map}
\end{equation}
Intuitively, this equation computes $\bm m$ by 1) mapping the current point $\bm x$ back to its initial location $\bm X = \psi(\bm x)$, 2) reading the initial impulse at $\bm X$, and 3) applying the deformation by multiplying with $\mathcal{T}^T$. The mathematical derivation can be found in the work by \citet{cortez1995impulse} (see Proposition 3).
The forward flow map $\phi$ and its Jacobian $\mathcal{F}$ are also leveraged in their system in a similar fashion. In particular, they evaluate:
\begin{equation}
    \bar{\bm m}(\bm X,0)=\mathcal{F}^T\,\bm m(\phi(\bm X),t),
    \label{eq:inverse_map}
\end{equation}
with $\bar{\bm m}(\bm X, 0)$ then being compared to ${\bm m}(\bm X, 0)$ for BFECC error compensation \cite{kim2006advections}. Their work's successful employment of $\phi$, $\psi$, $\mathcal{F}$ and $\mathcal{T}$ in a computer graphics simulation pipeline well-demonstrates the promise of flow map-based methods in building high-fidelity fluid solvers. 

\subsection{The Perfect Flow Map, and its Numerical Fallibilities}

Although the incorporation of $\mathcal{F}$ and $\mathcal{T}$ along with $\phi$ and $\psi$ has already led to a high level of simulation fidelity, the prevalent method used for solving these flow map variables \cite{nabizadeh2022covector, yin2023characteristic, qu2019efficient} is error-prone and invites reconsideration, as we will elaborate below.

\paragraph{The Perfect Flow Maps.}

In the absense of numerical inaccuracies, $\psi$, $\phi$, $\mathcal{F}$ and $\mathcal{T}$ should always have the following qualities:

\begin{remark}
\label{remark:roundtrip_psi}
A point undergoing, in sequence, a backward map $\psi$ and then a forward map $\phi$ should return to its original position. This also holds for the reverse direction. In other words,
\begin{equation}
    \begin{dcases}
        \bm X =\psi\circ\phi(\bm X), \\
        \bm x= \phi\circ\psi(\bm x).
    \end{dcases}
    \label{eq:roundtrip_psi}
\end{equation}
\end{remark}

\begin{remark}
\label{remark:roundtrip_psi}
A coordinate frame deformed, in sequence, by the backward map Jacobian $\mathcal{F}$ and then by the forward map Jacobian $\mathcal{T}$ should remain identical. This also holds for the reverse direction. In matrix notation,
\begin{equation}
    \begin{dcases}
        \bm I = \mathcal{F}\,\mathcal{T},\\
        \bm I = \mathcal{T}\,\mathcal{F}.
    \end{dcases}
    \label{eq:roundtrip_T}
\end{equation}
\end{remark}

\paragraph{Numerical Fallibilities}
Equations~\ref{eq:roundtrip_psi}~and ~\ref{eq:roundtrip_T} prescribe two consequential properties of flow maps which are challenging to satisfy numerically.
{\citet{qu2019efficient} and \citet{nabizadeh2022covector} employ} two separate numerical paradigms to evolve the backward and forward mappings. On one hand, the backward map $\psi$ is advected using a semi-Lagrangian-based method, which involves a diffusive grid interpolation at each step. On the other hand, the forward map $\phi$ is accurately solved by marching the map with a high-order Runge-Kutta method, which is void of interpolation errors. The asymmetry between both schemes leads to an \textit{inaccurate} backward mapping and an \textit{accurate} forward mapping, which are bound to diverge (become inconsistent) and fail to satisfy Equations~\ref{eq:roundtrip_psi}~and ~\ref{eq:roundtrip_T} numerically. 

\subsection{Alternative: Bidirectional March}
\label{subsec:bidirectional_march}
Driven by this observation, we propose a simple and natural alternative for solving $\phi$, $\psi$, $\mathcal{F}$ and $\mathcal{T}$: \textbf{bidirectional marching}. Since the forward and backward flow maps essentially describe the temporal integration of the velocity field forward and backward in time, we can adopt the same, interpolation-free, Runge-Kutta scheme for both the computation of ($\phi$, $\mathcal{F}$) and ($\psi$, $\mathcal{T}$), using $+\Delta t$ for the former and $-\Delta t$ for the latter. In particular, we solve for $\phi$ with Equation~\ref{eq:phi_def}, $\mathcal{F}$ with Equation~\ref{eq:F_evolution}; and solve for $\psi$ and $\mathcal{T}$ using the exact same procedures but with time reversed. 

\begin{figure}[t]
 \centering
 \includegraphics[width=.478\textwidth]{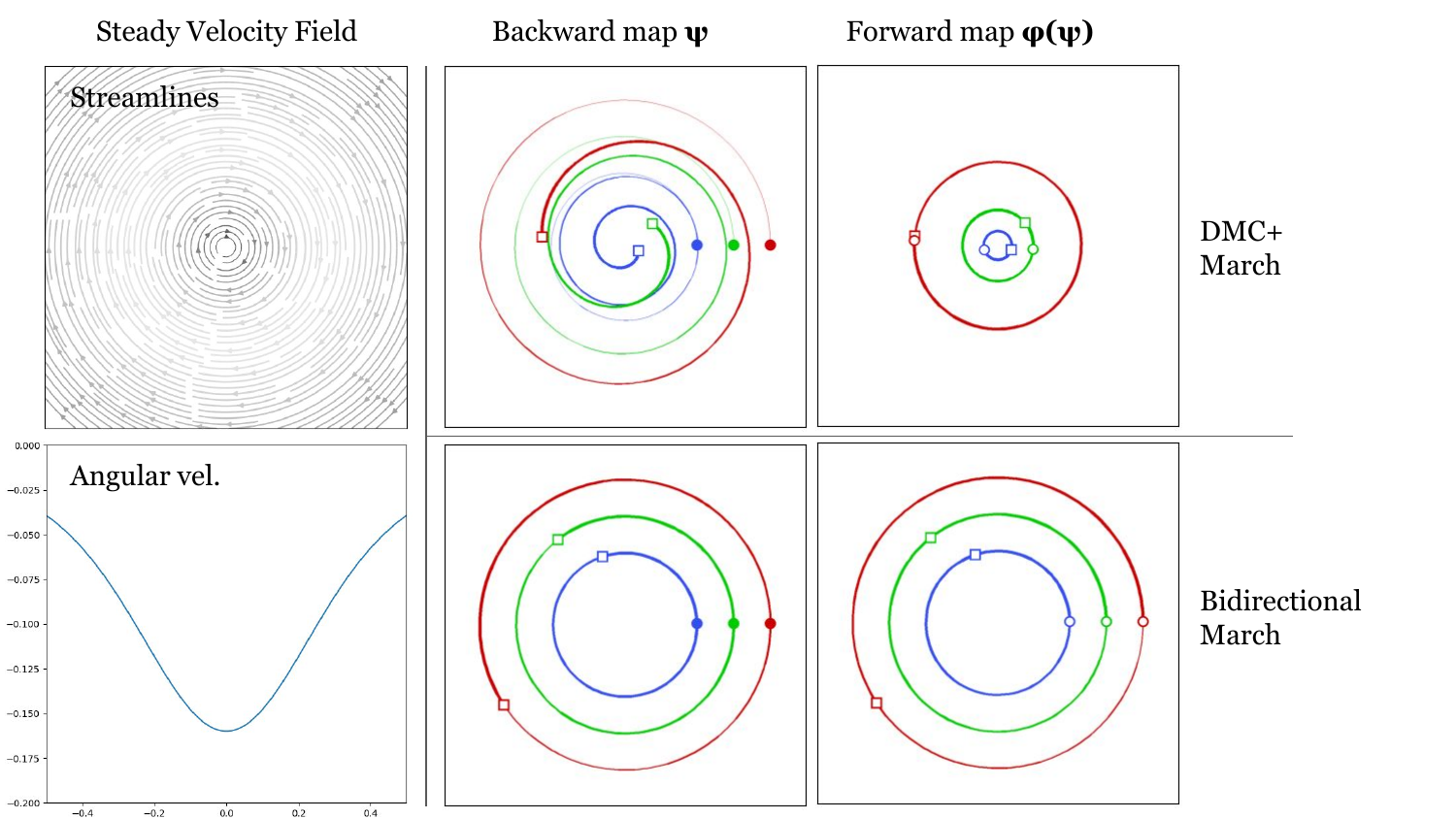}
 \caption{Under the steady velocity field shown on the left panel, a point transported by its flow map should 1) retain in its original orbit, and 2) return to its original position after the ``roundtrip'' of $\phi\circ\psi(x)$. This means that 1) the trajectories of the points should always be perfect circles, and 2) the solid circles should coincide with the hollow circles. It can be seen that \textit{Bidirectional March} satisfies these properties nicely, while \textit{DMC+March} yields significant errors.}
 \label{fig:advect_X_bad}
\end{figure}

\paragraph{A Motivational Experiment}
As shown in Figure~\ref{fig:advect_X_bad}, we conduct a concrete numerical experiment for both strategies \rev{to vividly illustrate the importance of symmetry in the computation of bidirectional flow maps}. We define a steady velocity field with a single vortex in the center, as shown on the left panel. The upper figure plots the streamlines, and the lower figure plots the angular velocity against the radial distance. We use this steady velocity field to construct flow map variables $\phi$, $\psi$, $\mathcal{F}$ and $\mathcal{T}$ for both methods, and compare their adherence to Equations~\ref{eq:roundtrip_psi}~and ~\ref{eq:roundtrip_T}.

The two figures on the top-right correspond to the solution adopted {by \citet{qu2019efficient} and \citet{nabizadeh2022covector}}, where a semi-Lagrangian-based scheme is used for solving $\psi$ (we use the dual-mesh
characteristics \cite{cho2018dual} for backtracking as suggested by \citet{qu2019efficient}), and the $4^\text{th}$ order Runge-Kutta scheme for solving $\phi$. Both $\mathcal{T}$ and $\mathcal{F}$ are computed from $\psi$ and $\phi$ using finite difference. The two figures on the bottom-right correspond to our proposed approach, where we solve for $\phi$ and $\psi$ with Equation~\ref{eq:phi_def}, $\mathcal{F}$ and $\mathcal{T}$ with Equation~\ref{eq:F_evolution} (top) --- all with $4^\text{th}$ order Runge-Kutta.

\begin{figure*}
 \centering
 \includegraphics[width=.99\textwidth]{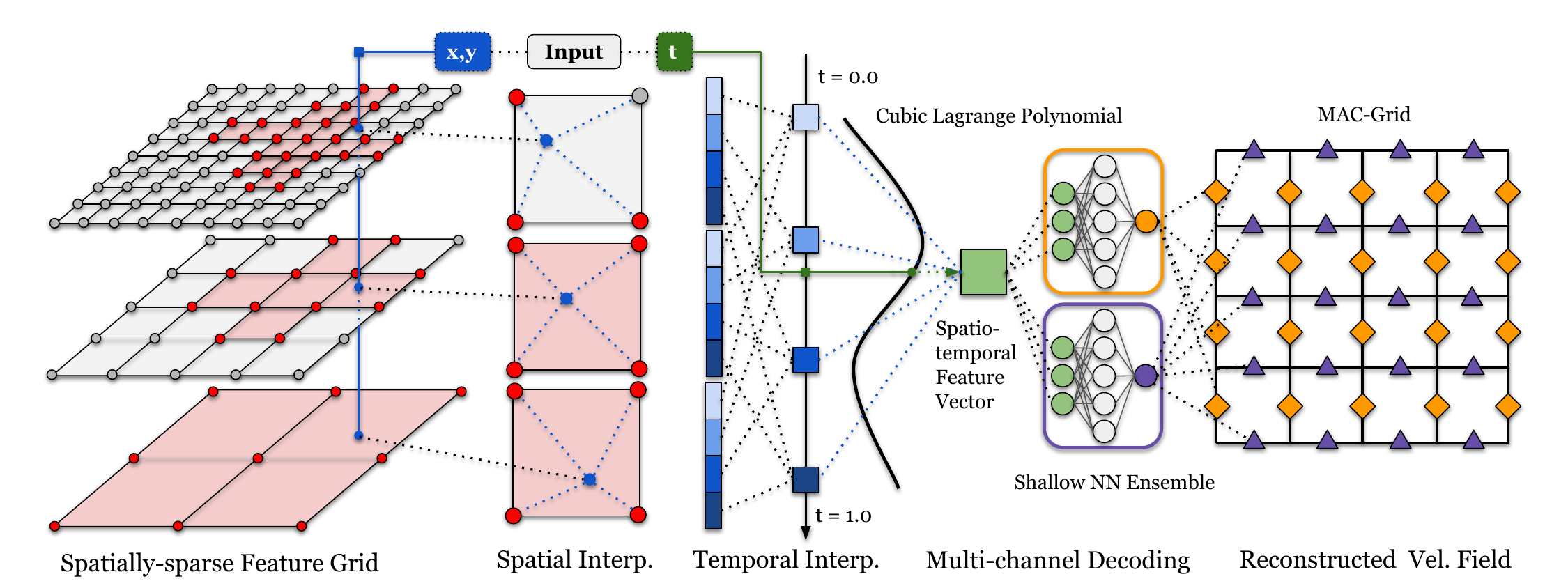}
 \caption{Illustration of our \inrfull (\inr) in 2D. In our \model simulation framework, we use \inr to represent a continuous spatiotemporal velocity field. To fetch the velocity given coordinates $(x,y,t)$, we first interpolate the multi-resolution, spatially sparse feature grid with $(x,y)$ to obtain one feature vector for each resolution (the left two columns). We reorganize these vectors into 4 temporal anchor vectors, and interpolate them with $t$ to obtain the final feature vector (middle column). Finally, we decode the feature vector with neural networks to get the velocity components (the right two columns).}
 \label{fig:INR_main}
\end{figure*}

The middle column depicts the backward mapping $\psi$, where the (input) current positions $\bm x$ are highlighted with \textbf{solid circles}, and the (backtracked) initial positions $\bm X = \psi(\bm x)$ are highlighted with \textbf{hollow squares}. The upper and lower plots in this column exhibit significantly different behaviors: the lower one traces out streamlines that are perfectly circular, whereas the upper one traces out streamlines that erroneously spiral inward, which are inconsistent with the ground-truth streamlines shown on the top-left.

The right column depicts the forward mapping $\phi$ originated from the (backtracked) initial positions $\bm X = \psi(\bm x)$. These are highlighted with \textbf{hollow squares}, and the roundtrip positions $\hat{\bm x} = \phi\circ\psi(\bm x)$ are highlighted with \textbf{hollow circles}. As prescribed by Equation~\ref{eq:roundtrip_psi}, $\hat{\bm x} = \phi\circ\psi(\bm x) = \bm x$, so the hollow circles ($\hat{\bm x}$) should in theory coincide with the solid circles (${\bm x}$). On the upper plot, this is clearly not the case with the presence of the radial drift that has previously occurred in the solving of $\psi$. On the lower plot, this requirement is well-fulfilled, due to that the computations of the backward and forward flow maps are mechanistically symmetric. 

The visual evidence is also supported numerically. The positional error of $\phi\circ\psi(\bm x) - \bm x$ is \textbf{0.3581} for \textit{DMC+March}, and \textbf{1.904E-06} for \textit{Bidirectional March}, showing that our proposed strategy better satisfies Equation~\ref{eq:roundtrip_psi} than the existing strategy \cite{qu2019efficient,nabizadeh2022covector} by 5 orders of magnitude. We further gauge both strategies' adherence to Equation~\ref{eq:roundtrip_T} by calculating the Frobenius norm of $\mathcal{F}\mathcal{T}-\mathbf{I}$. \textit{DMC+March} yields an error of \textbf{6.987}, while \textit{Bidirectional March} yields an error of \textbf{1.937E-3}, showing that our strategy better satisfies Equation~\ref{eq:roundtrip_T} by 3 orders of magnitude.

\subsection{Towards Perfect Flow Maps with INR}

\paragraph{Current Conundrum}
With the current flow map-based methods, the numerical inaccuracies of the flow maps undermine the simulation quality with the added 1) transportation error (\textit{e.g.}, warping the value from the wrong initial location), and 2) interpolation error by the requirement for frequent reinitializations.
To fulfill the full potential of low-diffusion, long-range flow map advection, a more accurate and symmetric method for computing the flow maps such as bidirectional marching would be desirable as it is proven to radically increase the numerical accuracy and consistency of the computed $\phi$, $\psi$, $\mathcal{F}$ and $\mathcal{T}$.
However, the key hindrance to the adoption of the bidirectional marching scheme is the intractable memory requirements for storing the spatiotemporal velocity buffer, which is necessitated for marching the backward variables $\psi$ and $\mathcal{T}$, as the solution from previous steps cannot be reused.

\paragraph{Our Solution}
Bridging such a conundrum faced by traditional fluid simulation with the state-of-the-art developments of fast-to-train, accurate, and memory-compact INRs in the machine learning and computer vision community, we design \textbf{\modelfull} (\model), an impulse and flow map-based fluid simulation method which implements the accurate {bidirectional marching} scheme for solving $\phi$, $\psi$, $\mathcal{F}$ and $\mathcal{T}$, at a small additional memory cost thanks to our novel hybrid INR structure: \textbf{\inrfull} (\inr), which is able to maintain long-range spatiotemporal velocity buffers at a smaller memory footprint than a single velocity field stored on a dense, uniform grid.

In Section~\ref{sec:method}, we first discuss the design and implementation of \inr. Then, in Section~\ref{sec:simulation}, we introduce our \model simulation algorithm which incorporates \inr at its core. Then, in Section~\ref{sec:validation}, we validate the efficacy of both \inr as a high-performance neural representation, and \model as a state-of-the-art simulation algorithm.

\setlength{\abovecaptionskip}{12pt}
\begin{figure*}[t]
 \centering
 \includegraphics[width=.99\textwidth]{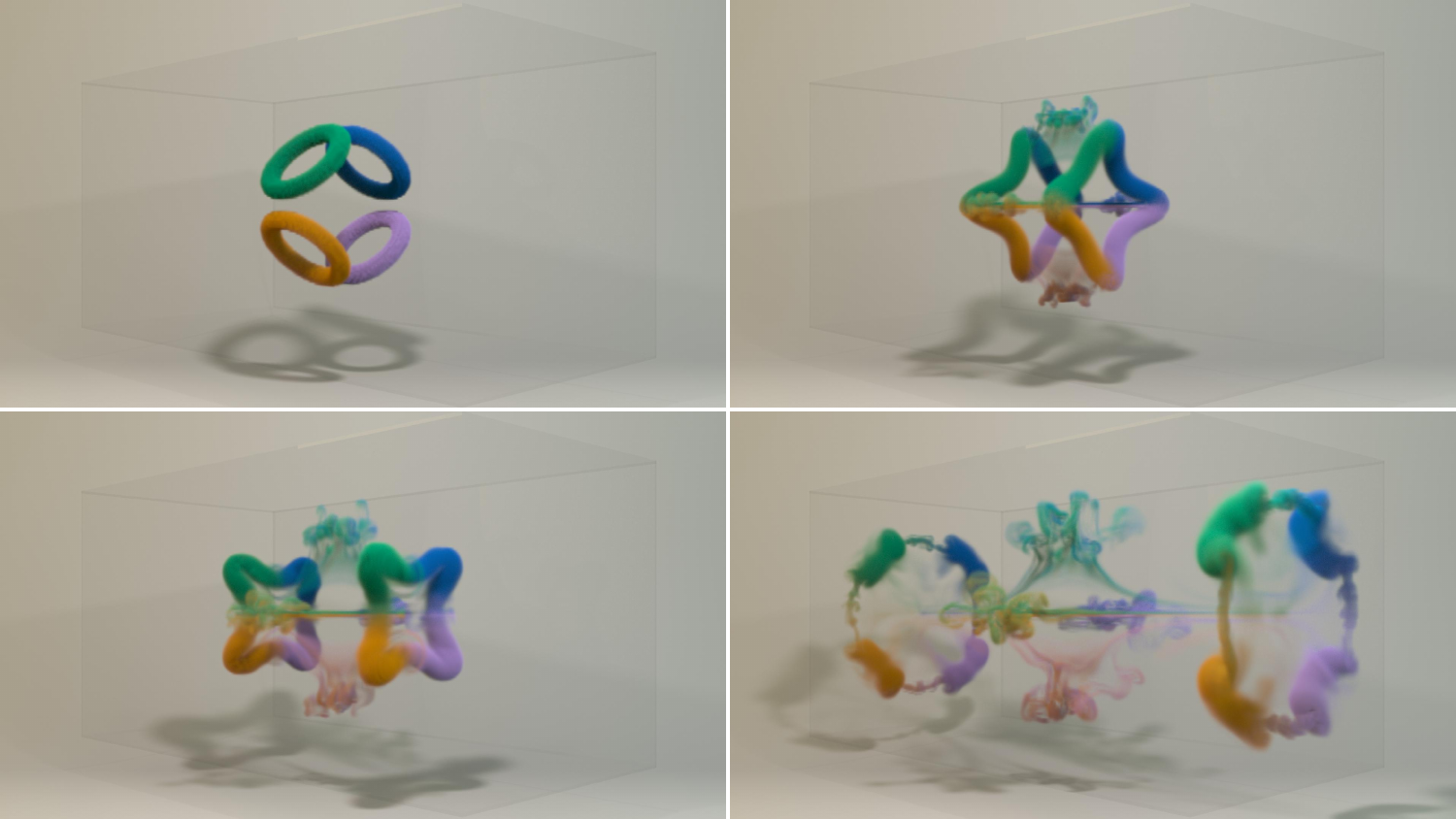}
 \caption{Collision and reconnection of four vortices each forming right angles with its neighbors. Upon collision, the four vortices reconnect to form two larger vortices shaped like four-pointed stars. They roam towards the left and right walls until each splits into four vortex tubes.}
 \label{fig:four_vorts}
\end{figure*}

\section{\inrfull}\label{sec:method}

In this section, we introduce \inrfull (\inr), which is the core component of our \modelfull simulator. As illustrated in Figure~\ref{fig:INR_main}, \inr is a novel, general-purpose hybrid INR for representing spatiotemporal signals. Specifically, in our application of fluid simulation, \inr represents a continuous spatiotemporal velocity field. \inr consists of 1) a multi-resolution, spatially sparse data structure that maintains trainable feature vectors, and 2) an ensemble of lightweight neural networks that decodes the feature vectors to velocity fields discretized on MAC grids. 

Reading Figure~\ref{fig:INR_main} from left to right: with input coordinates $(x,y,t)$, we first query the feature grid with spatial coordinates $(x,y)$ to fetch one feature vector at each resolution level. Each feature vector is split into 4 segments which are associated with 4 timestamps. The segments are concatenated across all resolution levels to form 4 temporal anchor vectors, which will be interpolated with $t$ to obtain the final feature vector. After that, we use our ensemble of shallow neural networks to decode this feature vector into the staggered $x$, $y$ (and $z$ in 3D) velocity components.

\subsection{Spatially Sparse Feature Grid}
\label{subsec:feature_grid}

As shown on the left of Figure~\ref{fig:INR_main}, we store feature vectors on a pyramid of multi-resolution regular grids, whose spatial sparsity is implemented with a bitmask. Our design is inspired by SPGrid~\cite{setaluri2014spgrid}, a shallow, sparse structure for physical simulation that is found to be more efficient and convenient than deep, tree-based structures. Unlike SPGrid which maintains a \textit{unique} discretization of the domain, our feature grids are \textit{overlapping}, therefore allowing a point in space to be included by cells on multiple levels and combine information from multiple spatial frequencies. This overlapping, multi-resolution design has been used to good effects by recent methods \cite{muller2022instant, takikawa2021neural}. 

\paragraph{Cell Activation}
Our spatial sparsity is explicitly controlled with a strategy similar to the octree division strategy proposed \rev{by \citet{ando2020practical}}. By default, the coarsest level is densely activated, and a cell on a \rev{finer} level is activated only when the \textit{complexity} of the flow at the region is found to exceed the characteristic \textit{expressiveness} of that level. The complexity is expressed with the sizing value $S$ computed from the Jacobian of the velocity field, as elaborated in Appendix~\ref{sec:sim_implementation}; and the characteristic expressiveness is given by \rev{$\sigma \cdot \frac{1}{dx}$ with \textit{activation threshold} $\sigma$ being a scalar hyperparameter}. Each cell on each level $l$ computes the max sizing $S_\text{max}$ among all the voxels it controls, and if $S_\text{max} > \sigma \cdot \frac{1}{dx_l}$, the cell becomes activated. As seen in Figure~\ref{fig:INR_main}, we activate the cell by activating all 4 \rev{(or 8 in 3D) geometric} nodes of the cell \rev{(the red dots)}. 
This process is carried out independently on all levels, and it guarantees that a voxel activated on one level is also activated on all the coarser levels. 

\begin{figure*}
 \centering
 \includegraphics[width=.99\textwidth]{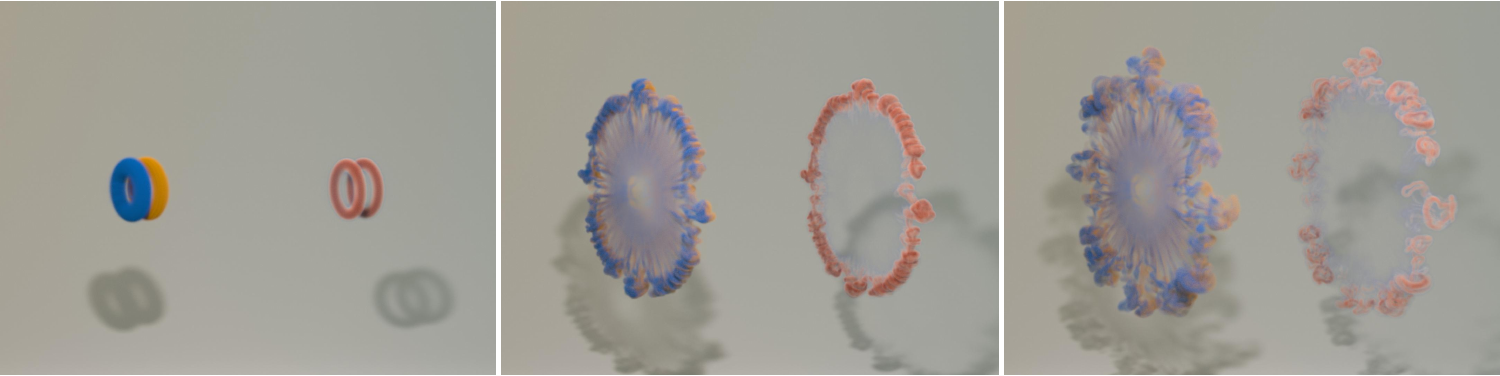}
 \caption{The evolution of a pair of vortex rings colliding head-on. Upon collision, the two rings stretch in the $yz$-plane while thinning in the $x$-direction. The structure then destabilizes and splits into a set of smaller vortex rings reminiscent of a Ferris wheel.}
 \label{fig:headon}
\end{figure*}

\paragraph{Feature Lookup}

On each level, we linearly interpolate a feature vector from the 4 (or 8 in 3D) enclosing nodes, and concatenate across all levels. If the interpolated location has no active nodes nearby, a zero vector is returned. If the enclosing nodes are a blend of active and inactive ones, we carry out the interpolation as usual while assuming the inactive ones contain zero vectors. \rev{Such an activation-agnostic lookup scheme requires no explicit case switching, and leads to the continuous transition between active and inactive regions, which improves the fitting quality. Quadratic interpolation has also been tested, but shows no significant improvement.}

\paragraph{Memory Management}

We use the \textit{bitmasked SNode} offered by Taichi \cite{hu2019taichi} to compose our data structure. Each bitmasked entry contains a dense block of $4^2$ (or $4^3$ in 3D) nodes to amortize the sparsity overhead. Each node stores in the AoS manner 1) a feature vector of length 8 (or 16 in 3D), and 2) a serialized index used for grid raveling. The gradients of the feature vectors are stored in a juxtaposed array, so the overall layout can be considered as SoAoS.

\paragraph{Comparison with Existing Methods} Our \inr is a collision-free analog of Instant~NGP \cite{muller2022instant}, as both methods employ multiple layers of multi-resolution, \rev{\textit{virtual}} dense grids, and concatenate the interpolated features. However, Instant~NGP saves memory by allowing multiple voxels to share the same parameters as assigned by a hashing function, and compete for dominance through their contributions to the loss function.  While this is reasonable for neural rendering applications, where surface cells dominate far-field and inner cells in the loss computation, it fails to disambiguate in our application where each velocity vector contributes equally to the loss. In comparison, we explicitly assign the parameters to the high-frequency regions, giving them exclusive access. As shown in the comprehensive numerical experiments in Subsection~\ref{subsec:inr_validation}, such explicit control acts to significantly reduce noise and leads to better fitting accuracy at lower parameter numbers and \rev{reduced training time}.
In this sense, \inr is also similar to NGLOD \cite{takikawa2021neural}, a method for fitting SDFs that also uses a collision-free sparse structure. In comparison, our overlapping grid-based design offers implementational simplicity and the alleviation of the restriction from surface octree to general-purpose, full-domain fields.
\rev{\inr is also comparable to NeuralVDB~\cite{kim2022neuralvdb} as both employ shallow structures rather than deep octrees for sparse representation, and both focus on \textit{compressing} volumetric input data instead of \textit{inferring} from images, although unlike \inr, NeuralVDB does not currently support temporal information. Both methods' design intents are also fundamentally different: NeuralVDB compresses readily-generated simulation data, whereas \inr is embedded in the simulation loop to facilitate high-quality simulation.}

\subsection{Temporal Dimension}
Our \inr handles the temporal dimension by interpolating multiple \textit{anchor} feature vectors in time, similar to the method \rev{by \citet{park2023temporal}}. Each feature vector looked up in space is split into 4 segments associated with timestamps $0$, $\frac{1}{3}$, $\frac{2}{3}$, and $1$ respectively. We use the cubic Lagrange polynomial to interpolate these vectors given any $t \in [0, 1]$. As validated in Subsection~\ref{subsec:inr_validation}, we find that cubic interpolation (instead of the linear interpolation found in existing methods) is consequential to the fitting accuracy.

\paragraph{Dynamic Timestamp Normalization} The interpolation scheme requires that the time of the stored sequence is within $[0, 1]$. For typical dynamic-NeRF applications, this is trivially fulfilled by normalizing the \textit{actual} timestamps by the total timelapse. In our case, this is challenging as the sequence is a stream of frames, each with a variable duration that is not  known in advance. In response, we devise \textit{dynamic timestamp normalization} to handle sequences with an indefinite number of frames or total timelapse. Essentially, we always assume that the current frames occupy the range $[0,1]$, and \rev{normalize the timestamps with the \textit{time multiplier} $\lambda$, which is the reciprocal of the timelapse-so-far}. As more of the sequence arrives, $\lambda$ decreases and the earlier frames are scaled down to accommodate the later frames. Not only does this lift the requirement for a predetermined timelapse, in Subsection~\ref{subsec:inr_validation}, we show that our strategy is more effective even when the total timelapse is available.

\subsection{Staggered Feature Decoder}
After the successful interpolation of feature vectors with spatiotemporal coordinates, the \inr will then reconstruct velocity fields on MAC grids using neural decoders. To obtain the staggered velocity components, we create one shallow neural network per spatial dimension. Based on the {shared} feature grid, for each spatial dimension, feature vectors are queried at its associated face centers and decoded with its proprietary neural network. Such a multi-branch design cannot be replaced with a single decoder that recovers the $x$, $y$ (and $z$ in 3D) components \textit{en masse}, because treating the staggered vector field as a scalar field leads to non-smooth signals.

\begin{figure*}
\centering
\begin{minipage}{.49\linewidth}
  \includegraphics[width=\linewidth]{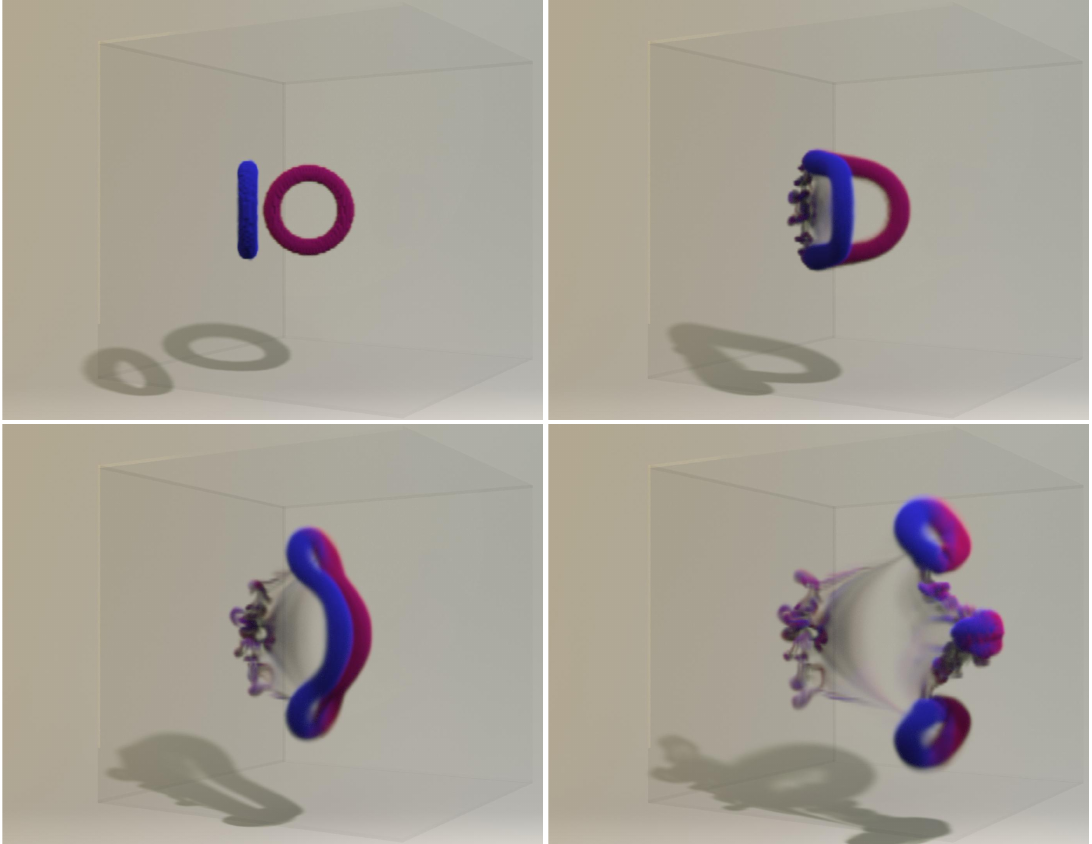}
  \captionof{figure}{The evolution of a pair of oblique vortex rings. The two vortices attach on the left side, undergo multiple topological changes, and eventually morph into three vortex rings.}
  \label{fig:oblique}
\end{minipage}
\hspace{.01\linewidth}
\begin{minipage}{.49\linewidth}
  \includegraphics[width=\linewidth]{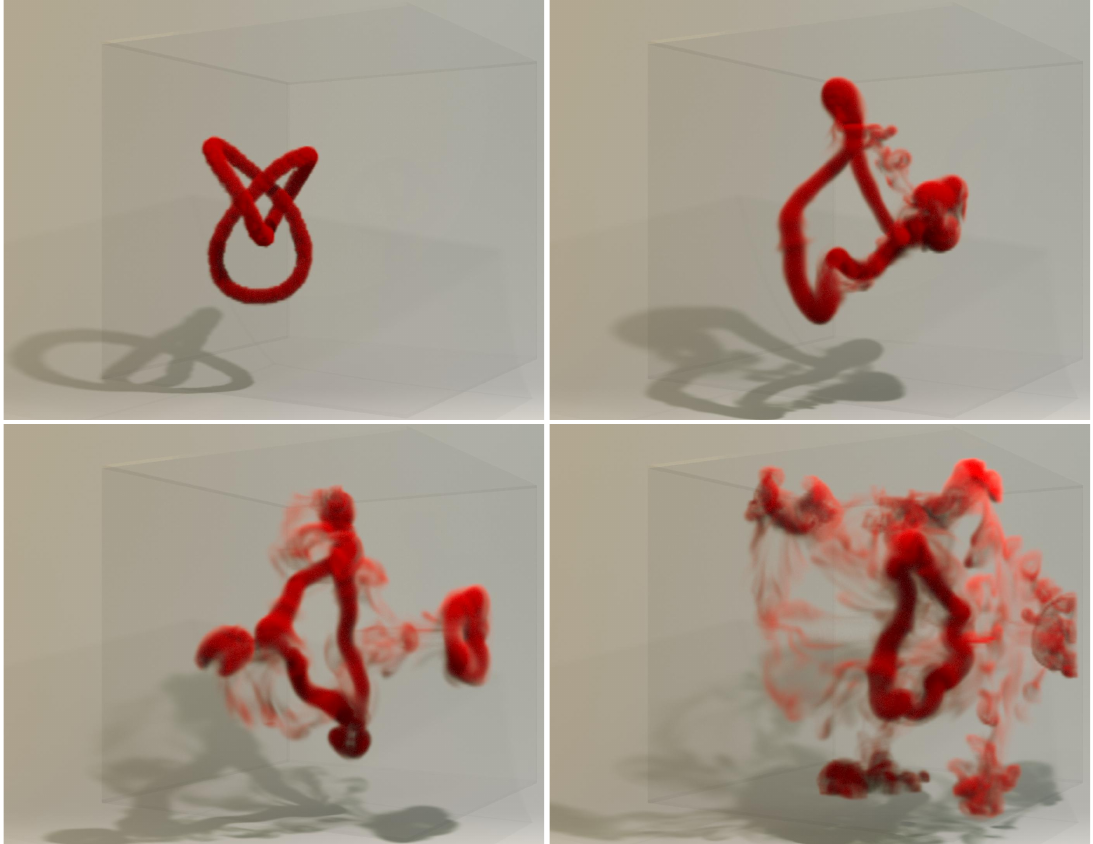}
  \captionof{figure}{The temporal evolution of a vortex trefoil knot. The knot deforms, self-collides, and breaks into two vortices of different sizes, corresponding to the experimental observation \cite{kleckner2013creation}.}
  \label{fig:trefoil}
\end{minipage}
\end{figure*}

\subsection{Training Scheme}
\label{subsec:training_scheme}
As an integral part of the \modelfull (\model) simulation loop, our \inr is trained at each simulation step to incorporate the latest update of the spatiotemporal velocity field produced by the simulation (elaborated in Section~\ref{sec:simulation}). During this training process, the input frame (the most recent velocity) must be properly stored \textit{without losing track of the previously stored frames}. However, the ground truth data for the prior frames are already lost. To address this, we maintain two copies of \inr: $\mathcal{N}$ and $\hat{\mathcal{N}}$, where the former is our main velocity buffer and the latter is an auxiliary buffer for memorizing the past. To better illustrate, suppose $\mathcal{N}$ has already stored the sequence $\bm{u}_1, \bm{u}_2, \bm{u}_3$.
When $\bm{u}_4$ arrives, we begin by hard-copying $\mathcal{N}$ to $\hat{\mathcal{N}}$. To update $\mathcal{N}$ for $\bm{u}_4$, we assemble the training data \textit{as if} we have $\bm{u}_1, \bm{u}_2, \bm{u}_3, \bm{u}_4$ in hand and sample all frames with equal priority --- but in practice, all the samples from $\bm{u}_1, \bm{u}_2, \bm{u}_3$ will be queried from $\hat{\mathcal{N}}$. \rev{We note that training recurrently in this way will indeed accumulate errors. But as shown in Figure~\ref{fig:fitting_plots} (left 4 plots), this is not significant in practice as the errors exhibit only a minor rising trend as the number of training sessions rises from 1 to 25.}

\paragraph{Data Sampling}
We train our \inr using mini-batches sampled from a discrete probability distribution $q$ computed as: \rev{
\begin{equation}
q_{I} = \alpha \cdot \min(\max\big(\sigma \cdot \frac{1}{dx_\text{C}}, S_{I}\big), \sigma \cdot \frac{1}{dx_\text{F}}),
\end{equation}
with $I$ denoting the 2 or 3D grid index, $dx_C$ and $dx_F$ the spacings of the coarsest and finest levels, and $\alpha$ the normalization factor to ensure that $q$ sums to 1. Such a distribution will bias the training data towards regions with greater flow complexity. The clipping functions are invoked to handle the edge cases (\textit{e.g., }without them, a constant field with $S = 0$ everywhere will leave $\alpha$ undefined).} More details regarding the sizing value $S$ can be found in Appendix~\ref{sec:sim_implementation}.

\paragraph{Encoder Growth} As discussed in Subsection~\ref{subsec:feature_grid}, the encoder is activated only for regions where the flow is deemed complex enough.
However, a region that is simple at the beginning of a sequence might become complex as the simulation proceeds. 
Hence, we \textit{grow} the encoder upon receiving a new frame by recomputing the sizing value $S$ and activating any additional cells as needed.

\paragraph{Reinitialization} 
As will be elaborated in Section~\ref{sec:simulation}, our simulation reinitializes every $n$ steps where \rev{$n$ is a hyperparameter whose selection should consider both \inr's fitting capacity and \model's accumulated errors from long-range flow map marching (further analysis can be found in Subsection~\ref{subsec:simulation_validation})}. 
At reinitialization, we uniformly randomly reinitialize the feature grid \rev{following \citet{muller2022instant}}, and leave the decoder networks untouched. Additionally, we also reinitialize the feature grid when the \rev{ratio of time multipliers $\frac{\lambda_{t-1}}{\lambda_t} > 1.33$} during {dynamic timestamp normalization}, 
because when the timestamp assignment for \textit{the same} frame changes significantly between two training sessions, the previously learned anchor vectors are often sub-optimal for initializing the current session.

\paragraph{Details} Each decoder is a shallow MLP with one hidden layer of width $64$. We use the ELU activation to avoid the dead ReLU issue~\citep{clevert2015fast}. We use AdamW~\citep{loshchilov2017decoupled} with $\beta = (0.9, 0.99)$ as the optimizer. The learning rate is set to $0.01$, and is scheduled to exponentially decay to $0.001$ at iteration $1500$. \rev{We employ the Mean Squared Error (MSE) as the loss function.}

\section{Simulation on \modelfull}\label{sec:simulation}

The \modelfull (\model) simulation algorithm built upon our \inr velocity buffer is outlined in Algorithm~\ref{alg:main}. 

At the start of a simulation, the \inr buffer will be \textit{randomly} initialized --- it serves purely as a data structure and hence does not need any pre-training.
Each simulation step starts by checking if the current step count is a multiple of $n$. If so, we reinitialize by setting the forward flow map $\phi$ and its Jacobian $\mathcal{F}$ to the identity mapping $\textbf{id}_\Omega$ and identity matrix $\bm{I}$ respectively, and the initial velocity $\bm{u}_{0}$ to the current velocity $\bm{u}$.
After that, we compute the current timestep $\Delta t_j$ ($j$ represents the step count since the last reinitialization), with $\bm{u}$ and the CFL condition. We maintain an array [$\Delta t_0, \dots, \Delta t_{n-1}$] to store the history of $\Delta t$.
Then, we compute the midpoint velocity $\bm u_\text{mid}$ using Algorithm~\ref{alg:midpoint}, and store it with the \inr $\mathcal{N}$. Doing so involves the encoder growth procedure and the utilization of the auxiliary buffer $\hat{\mathcal{N}}$, which have been discussed in Subsection~\ref{subsec:training_scheme}.

Once $\bm u_\text{mid}$ is properly stored in the \inr buffer $\mathcal{N}$, we compute $\phi$, $\psi$, $\mathcal{F}$ and $\mathcal{T}$ using $\mathcal{N}$. To do so, we reset $\psi$ and its Jacobian $\mathcal{T}$ to the identity map and identity matrix, and march with the spatiotemporal velocity field stored in $\mathcal{N}$ backward in time for a total of $j$ steps. For $\phi$ and $\mathcal{F}$, we reuse their solutions from the previous step and march for a single step forward in time. The marching algorithm is further elaborated in Subsection~\ref{subsec:flow_map_marching} and Algorithm~\ref{alg:RK4}.

Once $\phi$, $\psi$, $\mathcal{F}$ and $\mathcal{T}$ are obtained, we compute the advected velocity $\hat{\bm{u}}$ using the impulse-based fluid advection with BFECC. The detailed scheme is given in Algorithm~\ref{alg:bfecc}. 
If any external forces are present, their integrals along the fluid streamlines need to be computed and added to the velocity before projecting again by solving the Poisson equation.

\subsection{Bidirectional Flow Map Marching}\label{subsec:flow_map_marching}
As described in Subsection~\ref{subsec:bidirectional_march}, we solve for $\phi$ and $\psi$ by marching Equation~\ref{eq:phi_def} forward and backward in time; and solve for $\mathcal{F}$ and $\mathcal{T}$ by marching Equation~\ref{eq:F_evolution} (top) forward and backward in time.

All four quantities are temporally integrated with our custom $4^\text{th}$ order Runge-Kutta (RK4) integration scheme, outlined in Algorithm~\ref{alg:RK4}.
The algorithm uses $\phi$ and $\mathcal{F}$ for notation, but applies directly to $\psi$ and $\mathcal{T}$ by reversing time. 
We note that, to evolve $\mathcal{F}$ and $\mathcal{T}$ using the RK4 scheme, the values $\frac{\partial \mathcal{F}}{\partial t}$ and $\frac{\partial \mathcal{T}}{\partial t}$ need to be estimated at $t + 0.5\Delta t$ and $t + \Delta t$, which require the estimations of $\phi$ and $\psi$ at these times. We manage this by evolving the maps and the Jacobians in an \textit{interleaved} manner, so that the estimations of $\phi$ and $\psi$ at $t + 0.5\Delta t$ and $t + \Delta t$ can be recycled for updating $\mathcal{F}$ and $\mathcal{T}$.

\subsection{Midpoint Method}
As outlined in Algorithm~\ref{alg:midpoint}, we adopt the second-order, midpoint method to effectively reduce the truncation error of the temporal integration. To do so, we reset $\psi$ and $\mathcal{T}$ to identity, perform a single RK4 backtrack with the current velocity $\bm{u}$ and $0.5\cdot\Delta t$, and carry out the impulse-based advection to obtain $\bm{u}_\text{mid}$. We empirically find that, for estimating $\bm{u}_\text{mid}$, neither the BFECC error compensation nor the long-range flow map makes a significant difference to the simulation results, hence they are ablated for the estimation of $\bm{u}_\text{mid}$.

\begin{algorithm}[t]
\caption{Neural Flow Map Simulation}
\label{alg:main}
\begin{flushleft}
        \textbf{Initialize:} $\bm{u}$ to initial velocity; $\mathcal{N}$, $\hat{\mathcal{N}}$ with random weights
\end{flushleft}
\begin{algorithmic}[1]
\For{$i$ in total steps}
\State $j \gets i \Mod n$;
\If{$j$ = 0}
\State $\phi \gets \textbf{id}_\Omega$, $\mathcal{F} \gets \bm{I}$;
\State $\bm{u}_0 \gets \bm{u}$;
\State Randomly initialize $\mathcal{N}$'s feature vectors;
\EndIf
\State Compute $\Delta t_j$ with $\bm{u}$ and the CFL number;
\State Estimate midpoint velocity $\bm{u}_\text{mid}$ according to Alg. \ref{alg:midpoint};
\If{$i+1 \Mod n \neq$ 0}
\State Compute sizing field $S$ with $\bm{u}_\text{mid}$;
\State Grow $\mathcal{N}$'s feature grid with $S$;
\State Train $\mathcal{N}$ with $\bm{u}_\text{mid}$ and $\hat{\mathcal{N}}$;
\State $\hat{\mathcal{N}} \gets \mathcal{N}$;
\EndIf
\State $\psi \gets \textbf{id}_\Omega$, $\mathcal{T} \gets \bm{I}$;
\State March $\psi, \mathcal{T}$ with Alg.~\ref{alg:RK4}, using $\bm{u}_\text{mid}$ and $-\Delta t_j$;
\For {$l$ in $j-1 \dots 0$}
\State Query $\bm{u}_{\text{mid}, l}$ from $\mathcal{N}$ using time $\sum_{c=0}^{l-1}\Delta t_c + 0.5 \Delta t_l$;
\State March $\psi, \mathcal{T}$ with Alg.~\ref{alg:RK4}, using $\bm{u}_{\text{mid}, l}$ and $-\Delta t_l$;
\EndFor{}
\State March $\phi, \mathcal{F}$ according to Alg.~\ref{alg:RK4}, using $\bm{u}_\text{mid}$ and $\Delta t_j$;
\State Reconstruct $\bm{u}$ with $(\bm{u}_0, \psi, \mathcal{T}, \phi, \mathcal{F})$ as in Alg. \ref{alg:bfecc};
\If{use external force}
\State $\hat{\bm{u}} \gets \bm{u} + \int_0^t\bm{f}_\text{ext}(\psi(\bm x, \tau), \tau)d\tau$;
\State $\bm{u} \gets \textbf{Poisson}
(\hat{\bm{u}})$;
\EndIf
\EndFor{}
\end{algorithmic}
\end{algorithm}

\subsection{Error-compensated Impulse Advection}

As outlined in Algorithm~\ref{alg:bfecc}, once we have successfully solved for the quantities $\phi$, $\psi$, $\mathcal{F}$ and $\mathcal{T}$, we transport the initial velocity $\bm{u}_0$ to find the updated velocity $\bm{u}$ using the impulse-based scheme \cite{cortez1995impulse, nabizadeh2022covector}, which is used in combination with the BFECC error compensation \cite{kim2006advections, qu2019efficient}.

\rev{Additional implementation details of our \model simulation method can be found in Appendix~\ref{sec:sim_implementation}.}

\section{Validation}\label{sec:validation}

We validate the effectiveness of \model in two parts. First, we conduct a thorough comparison of \inr with existing INR benchmarks in terms of memory efficiency, training speed, and fitting accuracy. Then, we validate the advantage of employing \inr for simulation, by comparing \model quantitatively and qualitatively with traditional simulation methods, showcasing improved adherence to analytical solutions, energy preservation, and phenomenological fidelity.

\begin{figure}
 \centering
 \includegraphics[width=.47\textwidth]{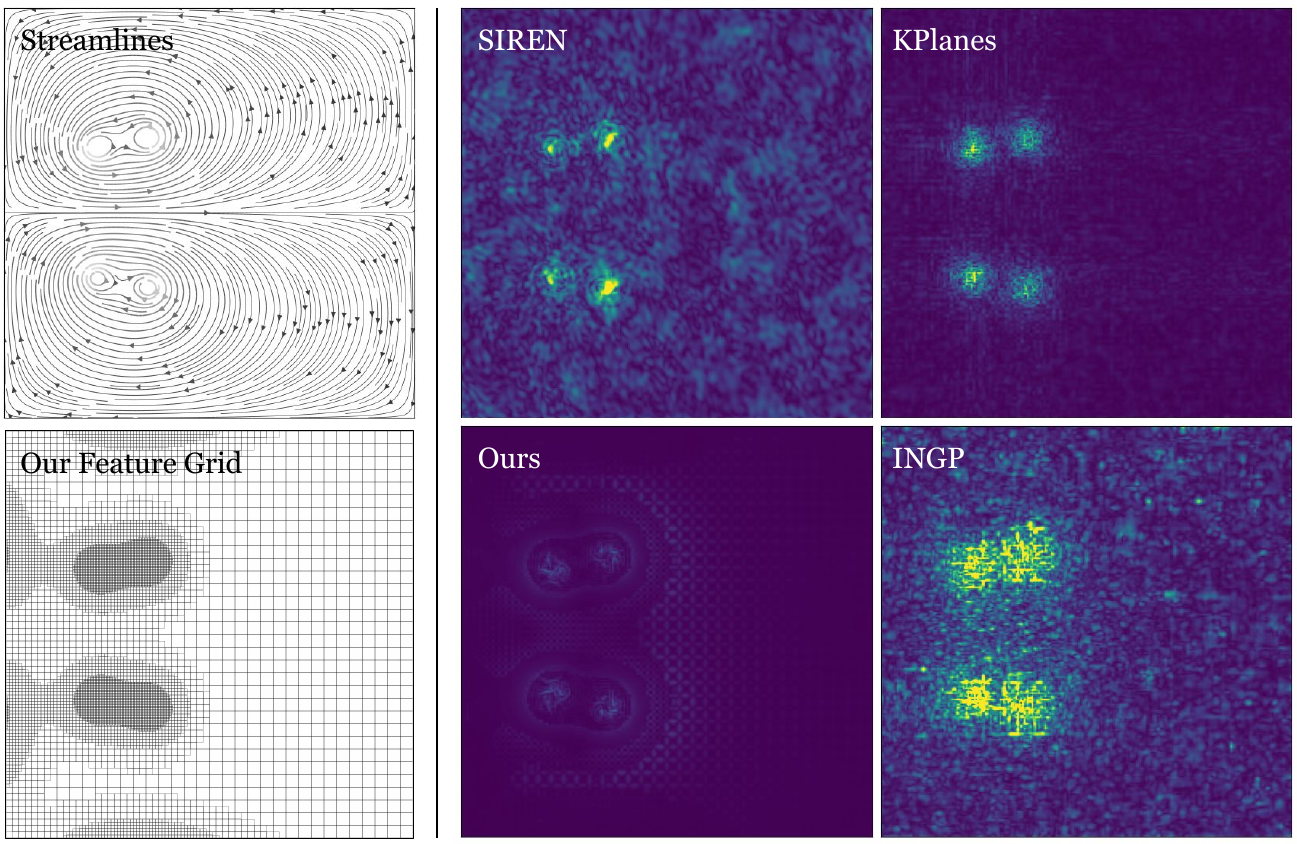}
 \caption{Fitting error in 2D. The benchmarks are noisy around the vortical regions; our method achieves high fitting accuracy in these regions at a smaller memory footprint due to its explicitly managed spatial sparsity.}
 \label{fig:2D_error}
\end{figure}

\subsection{Validation of \inr}
\label{subsec:inr_validation}

We compare \inr to three benchmarks: Instant NGP (INGP) \cite{muller2022instant}, KPlanes \cite{fridovich2023k}, and SIREN \cite{sitzmann2020implicit}, in fitting spatiotemporal fields in both 2D and 3D (spatial-wise). These benchmarks are carefully selected to cover a wide spectrum of INR paradigms. INGP and KPlanes are both hybrid INRs combining feature vectors stored on traditional data structures with small neural networks, a design adopted by our method. Between them, INGP achieves data compression by leveraging spatial sparsity, as does our approach, whereas KPlanes is spatially dense and achieves compression by spatial decomposition. SIREN, on the other hand, is a pure INR that is fully implicit featuring much larger neural networks. \rev{Detailed descriptions of each benchmark along with our testing methodology can be found in Appendix~\ref{sec:experimental_setup}}.

\setlength{\abovecaptionskip}{12pt}
\begin{figure*}[t]
 \centering
 \includegraphics[width=.99\textwidth]{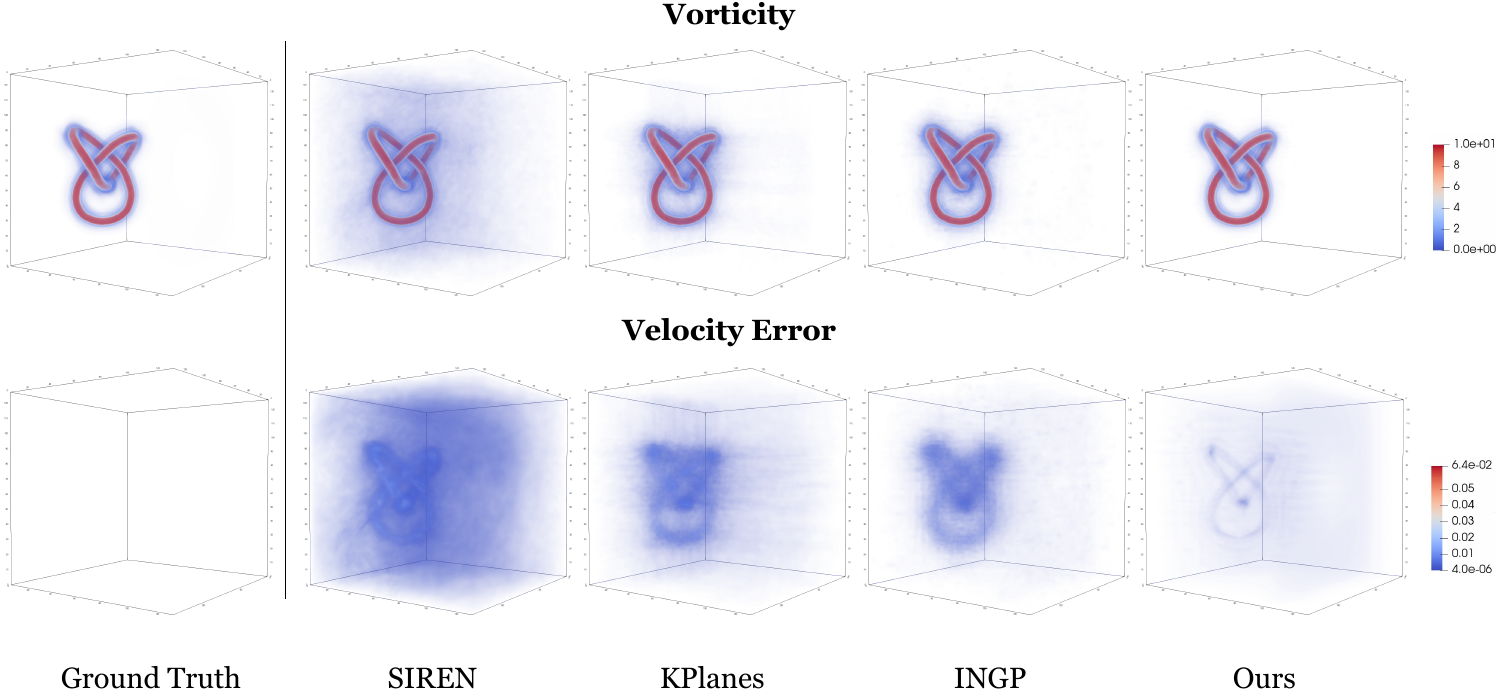}
 \caption{Visualization of the fitting quality of a spatiotemporal sequence in 3D. Displayed is the $8^\text{th}$ frame from the $24^\text{th}$ training session. The top row shows the vorticity field. It can be seen that our method yields the highest-quality compression with the lowest amount of smearing around the vortex tubes. The bottom row shows the velocity error compared to the ground truth, which confirms the improved accuracy of our method, especially near the vortical regions.}
 \label{fig:3D_error}
\end{figure*}

\begin{figure*}[t]
 \centering
 \includegraphics[width=.99\textwidth]{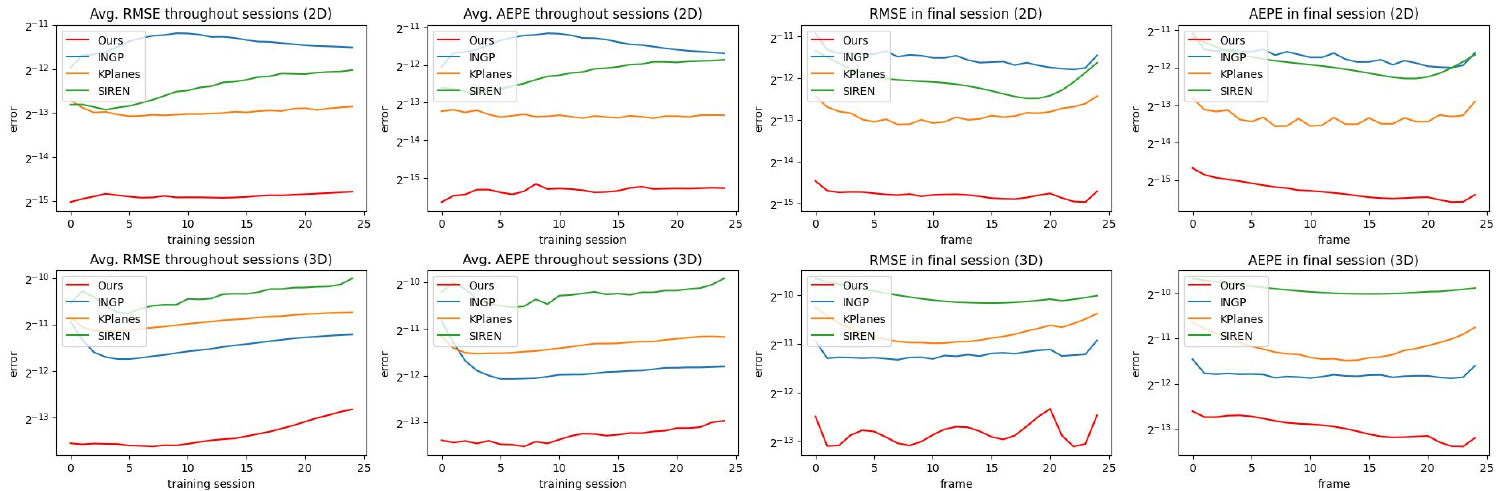}
 \caption{The fitting errors (RMSE and AEPE) for experiments in 2D (top) and 3D (bottom). Our method consistently yields the lowest fitting error in comparison to existing methods.}
 \label{fig:fitting_plots}
\end{figure*}

The testing results are presented in Table~\ref{tab: INR_validation_table} and Figure~\ref{fig:fitting_plots}. On the left of Table~\ref{tab: INR_validation_table}, the computational cost of the 2D and 3D versions of all 4 methods are given. On the right, the averaged errors are given. Since the errors are {natively} computed on the MAC grid, we report two errors: a scalar (component-wise) error with the Root Mean Squared Error (RMSE), and a vector error with the Average End Point Error (AEPE). The velocity vectors are reconstructed by aggregating the face-centered velocity components at the cell centers. For each metric, we present three numbers: the initial error, the final error, and the averaged error. To elucidate the difference between these errors, we remind that each test consists of 25 \textit{training sessions}. The initial session learns a single, static frame, while the final session learns the entire sequence of 25 frames. Hence the initial error isolates the system's capacity to resolve spatial details, while the final error gauges its capacity to handle the long-range temporal evolution. The average error measures the system's consistency throughout the 25 training sessions. 

The left of Table~\ref{tab: INR_validation_table} shows that, in 2D, \inr is the most memory efficient and only trails INGP in terms of time cost by less than 10\%. In 3D, \inr is the most time efficient and only trails SIREN in terms of memory cost, albeit for a favorable tradeoff since SIREN trains about $6\times$ slower. 
The right of Table~\ref{tab: INR_validation_table} shows that \inr has a clear advantage in terms of fitting accuracy. For instance, in 2D, our method reduces the average RMSE by 73.7\% from the best benchmark (KPlanes) and 91.1\% from the worst benchmark (INGP). In 3D, our method reduces the error by 73.5\% from the best benchmark (INGP) and 87.3\% from the worst benchmark (SIREN). Similarly, in Figure~\ref{fig:fitting_plots}, it can be observed that our method consistently outperforms the benchmarks in all metrics.

\paragraph{Discussion}
The reason behind our method's effectiveness can be probed with Figure~\ref{fig:2D_error}, in which the ground truth velocity is plotted on the top left, and the fitting errors of all four methods are plotted on the right. The feature grid discretization of our method is illustrated on the lower left. We can see that for all the benchmarks, the error is concentrated in the vortical regions with large velocity gradients. These methods either lack the ability to adapt the DoFs towards these regions (KPlanes), or rely on black-box schemes (INGP, SIREN) which underperform for our application. In comparison, our method explicitly assigns its DoFs to the regions where they are the most needed, according to the domain-specific prior knowledge encoded in the sizing function (Equation~\ref{eq:sizing} of Appendix~\ref{sec:sim_implementation}). Despite \inr using the fewest parameters, the lower left of Figure~\ref{fig:2D_error} shows that most of the parameters are clustered to densely cover the vortical regions, which leads to a much greater local effective resolution and therefore much smaller fitting errors. 

The analogous 3D error visualization is given in Figure~\ref{fig:3D_error}, in which the vorticity fields computed from the fitted velocity fields are depicted on the top row, and the fitting errors are visualized on the bottom row. It can be observed that \inr yields the highest-fidelity fitting result by a considerable margin, sharply carving out high-frequency details near the trefoil knot, while the benchmarks are visibly diffused around the knot, a sign of insufficient storage resolution. Our method achieves this by explicitly assigning most of its parameters to express the knot structure, so that the effective resolution is largely increased at a manageable cost.

\setlength{\abovecaptionskip}{7pt}
\begin{table*}[t]
\centering\small
\begin{tabularx}{\textwidth}{Y | Y | Y || Y | Y | Y | Y |Y| Y}
\hlineB{3}
\multicolumn{3}{c||}{\textbf{Computational Cost}} & \multicolumn{6}{c}{\textbf{Performance}}\\
\hlineB{2.5}
& Num. Params. & Time (s) & RMSE (init.) & AEPE (init.) & RMSE (avg.) & AEPE (avg.) & RMSE (final) & AEPE (final) \\
\hlineB{2.5}
\multicolumn{9}{c}{\textbf{XY + T}}\\
\hlineB{2.5}
INGP & 90,562 & \textbf{0.49} & 24.91E-5 & 23.66E-5 & 36.89E-5 & 35.82E-5 & 34.33E-5&30.25E-5\\
\hlineB{2}
KPlanes & 105,262 & 0.64 & 15.00E-5 & 10.42E-5 & 12.55E-5 & 9.659E-5 & 13.51E-5& 9.679E-5\\
\hlineB{2}
SIREN & 98,822 & 1.19 & 13.95E-5 & 16.11E-5 & 18.53E-5 & 21.16E-5 & 24.02E-5 & 26.89E-5\\
\hlineB{2}
Ours & \textbf{83,138 - 87,522} & 0.53 & \textbf{2.999E-5} & \textbf{1.960E-5} & \textbf{3.299E-5} & \textbf{2.439E-5} & \textbf{3.533E-5} & \textbf{2.536E-5} \\
\hlineB{2.5}
\multicolumn{9}{c}{\textbf{XYZ + T}}\\
\hlineB{2.5}
INGP & 2,148,147 & 4.26 & 51.36E-5 & 54.98E-5 &36.12E-5&27.59E-5&42.13E-5&27.97E-5\\
\hlineB{2}
KPlanes & 2,335,107 & 2.99 & 54.43E-5 & 43.75E-5 &51.48E-5&38.66E-5&58.60E-5&43.42E-5\\
\hlineB{2}
SIREN & \textbf{462,595} & 11.60 & 67.10E-5 & 84.89E-5 &75.06E-5&82.54E-5&96.94E-5&103.6E-5\\
\hlineB{2}
Ours & {1,969,779 - 2,129,891} & \textbf{1.94} & \textbf{8.356E-5} & \textbf{9.323E-5} & \textbf{9.554E-5} & \textbf{10.01E-5} & \textbf{13.84E-5} & \textbf{12.46E-5} \\
\hlineB{3}
\end{tabularx}
\caption{Errors of our method compared to those of the three benchmarks in spatiotemporal signal fitting in 2D and 3D. \rev{The time reported reflects the training time per 100 training iterations.} Our method consistently yields the lowest fitting error at a highly competitive memory and time cost.}
\label{tab: INR_validation_table}
\vspace{0pt}
\end{table*}

In summary, experimental evidence suggests that our \inr representation is generally advantageous for cases where:
\begin{enumerate}
    \item The accuracy of compression is important, \textit{e.g.}, in scientific computing;
    \item The memory budget is constrained and needs to be efficiently managed;
    \item The spatial sparsity is significant and domain-specific knowledge is available.
\end{enumerate}

Therefore, suitable scenarios for \inr \textit{include} high-fidelity fluid simulations for computer graphics and computational fluid dynamics, but they are also not confined to this application.

\begin{figure}[t]
 \centering
 \includegraphics[width=.47\textwidth]{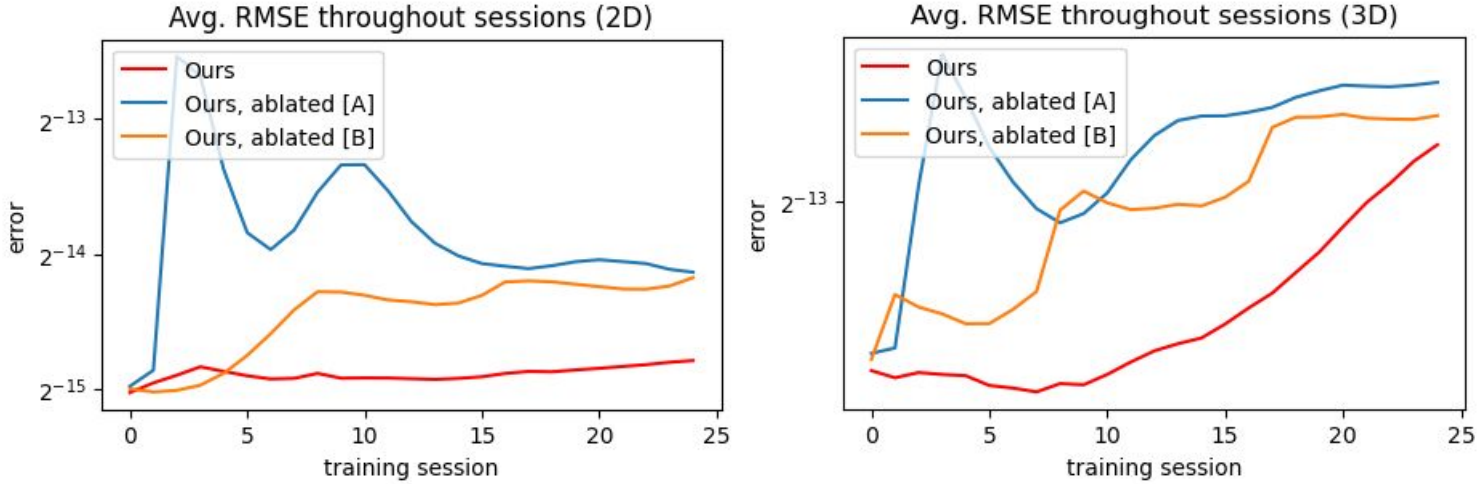}
 \caption{Errors of the ablation tests corresponding to Table~\ref{tab: ablation_table}.}
 \label{fig:ablation_plots}
\end{figure}

\paragraph{Ablation Studies}
As shown in Table~\ref{tab: ablation_table} and Figure~\ref{fig:ablation_plots}, we conduct 2D and 3D comparisons with two ablated versions: [A] without dynamic timestamp normalization (instead, normalize full timelapse to $[0,1]$), and [B] with linear time interpolation instead of the cubic one. To facilitate [A], we create test sequences with a fixed per-step duration, so that the total timelapse is known ahead of time.
It can be seen that both ablated versions are less effective than the full version, which validates that our proposed techniques are well-motivated.

\paragraph{Influence of Encoder's Size}
\rev{
As shown in Figure~\ref{fig:hyperparameters_ablation} (right), we analyze the influence of the activation threshold $\sigma$ and the encoder's effective resolution, which together control the encoder's number of parameters. We perform 2 experiment sets: 1) fixing $\sigma = 0.01$ with varying \textit{base} resolutions $(8, 32)$, $(16, 64)$, $(32, 128)$, and $(64, 256)$ (with 4 levels of refinement), and 2) fixing base resolution $(32, 128)$ with varying $\sigma =$ 0.01, 0.02, 0.03, 0.04, and 0.05. For 1), a higher resolution leads to more parameters, and for 2), a lower $\sigma$ leads to more parameters. For both tests, we observe that the accuracy quickly saturates once the encoder reaches a certain size, and a larger model can even worsen the fitting accuracy, due to our limited training iterations. 
}

\subsection{Validation of \model Simulation}
\label{subsec:simulation_validation}
In this section, we validate the efficacy of our \model neural simulation method in comparison with existing benchmarks. We begin with a simple, 2D steady flow to compare our method to the benchmarks in retaining the steady state. We move on to complex 3D scenarios without analytical solutions and compare in terms of the conservation of energy. Finally, we qualitatively compare our method to the benchmarks in terms of visual intricacy and the recreation of real-world phenomena.

\begin{table}[t]
\centering\small
\begin{tabularx}{0.47 \textwidth}{Y | Y | Y | Y | Y}
\hlineB{3}
& 2D RMSE (avg.) & 2D RMSE (final) & 3D RMSE (avg.) & 3D RMSE (final)\\
\hlineB{2.5}
Ours & \textbf{3.299E-5} & \textbf{3.533E-5} & \textbf{9.554E-5} & \textbf{13.84E-5}\\
\hlineB{2}
Ablated [A] & 7.258E-5 & 5.562E-5 & 13.94E-5 & 15.91E-5\\
\hlineB{2}
Ablated [B] & 4.555E-5 & 5.408E-5 & 12.12E-5 & 14.78E-5\\
\hlineB{3}
\end{tabularx}
\vspace{5pt}
\caption{Time-averaged ablation results. The reduced errors of our full method validate that our design choices are well-motivated.}
\label{tab: ablation_table}
\vspace{-8pt}
\end{table}

\begin{figure}[t]

\includegraphics[width=0.472\linewidth]{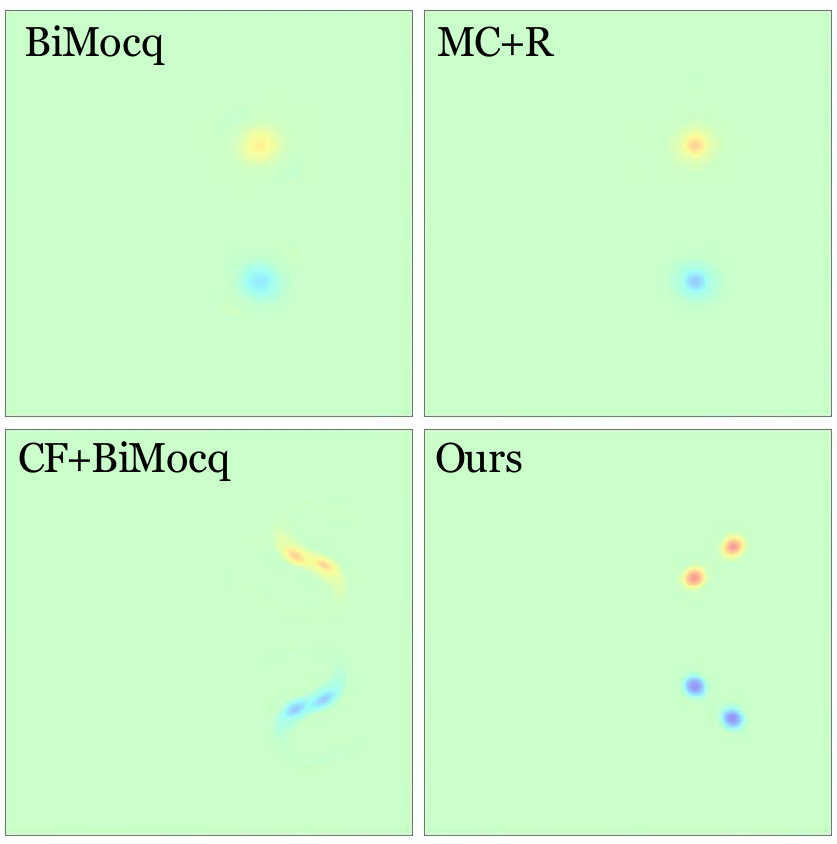} 
\includegraphics[width=0.518\linewidth]{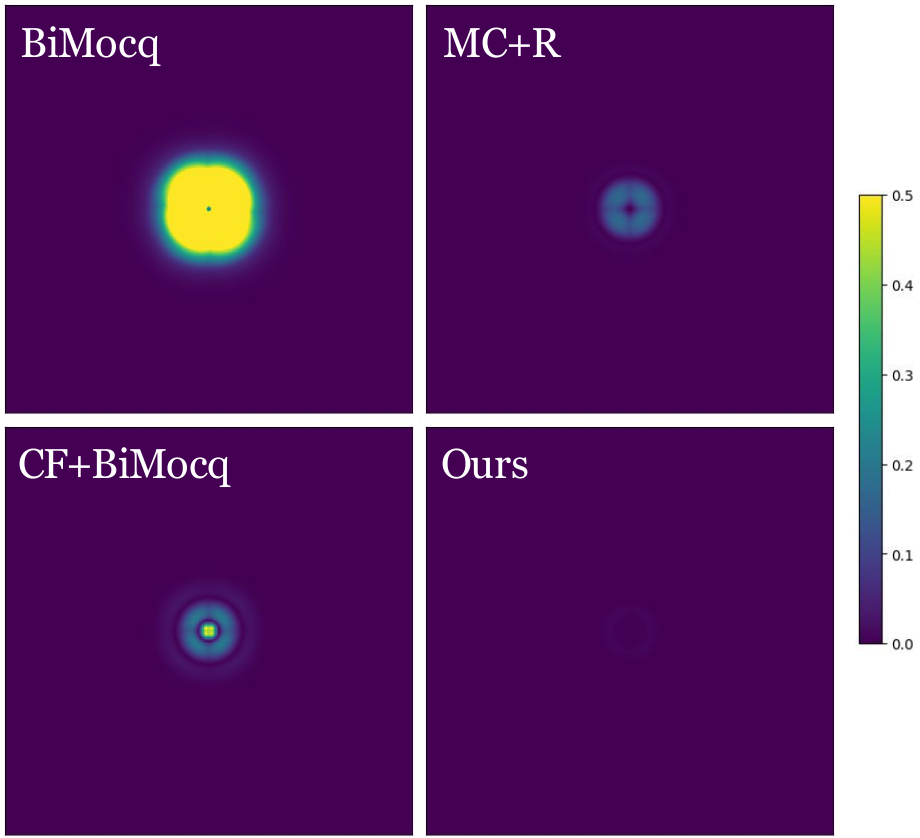}

\caption{Left: leapfrogging vortices in 2D. Our method best preserves the vortical structures. Right: absolute errors \textit{w.r.t.} the steady-state field after 1300 frames. The mean absolute errors are 1.589E-4, 21.56E-4, 30.92E-4, and 488.9E-4 for \model (ours), MC+R, CF+BiMocq, and BiMocq respectively.}
\label{fig:two_2d_comparisons}
\end{figure}

\begin{figure}[h]

 \includegraphics[width=0.48\linewidth]{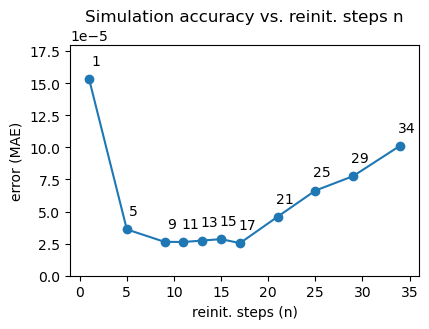} 
\includegraphics[width=0.51\linewidth]{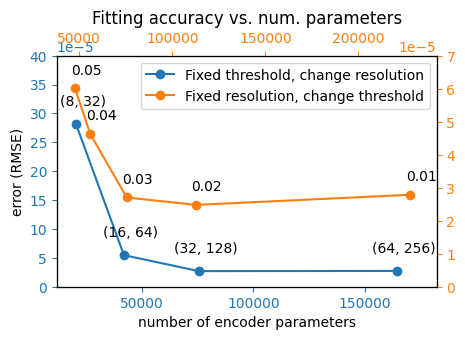}
 \caption{\rev{Left: simulation errors from different choices of $n$. Right: fitting errors for different encoder sizes.}}
\vspace{-1pt}
\label{fig:hyperparameters_ablation}
\vspace{15pt}
\includegraphics[width=0.47\linewidth]{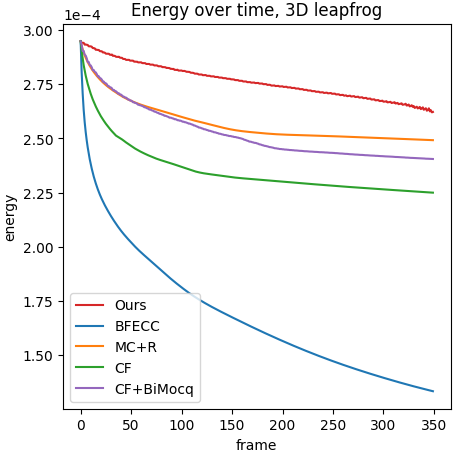} 
\includegraphics[width=0.465\linewidth]{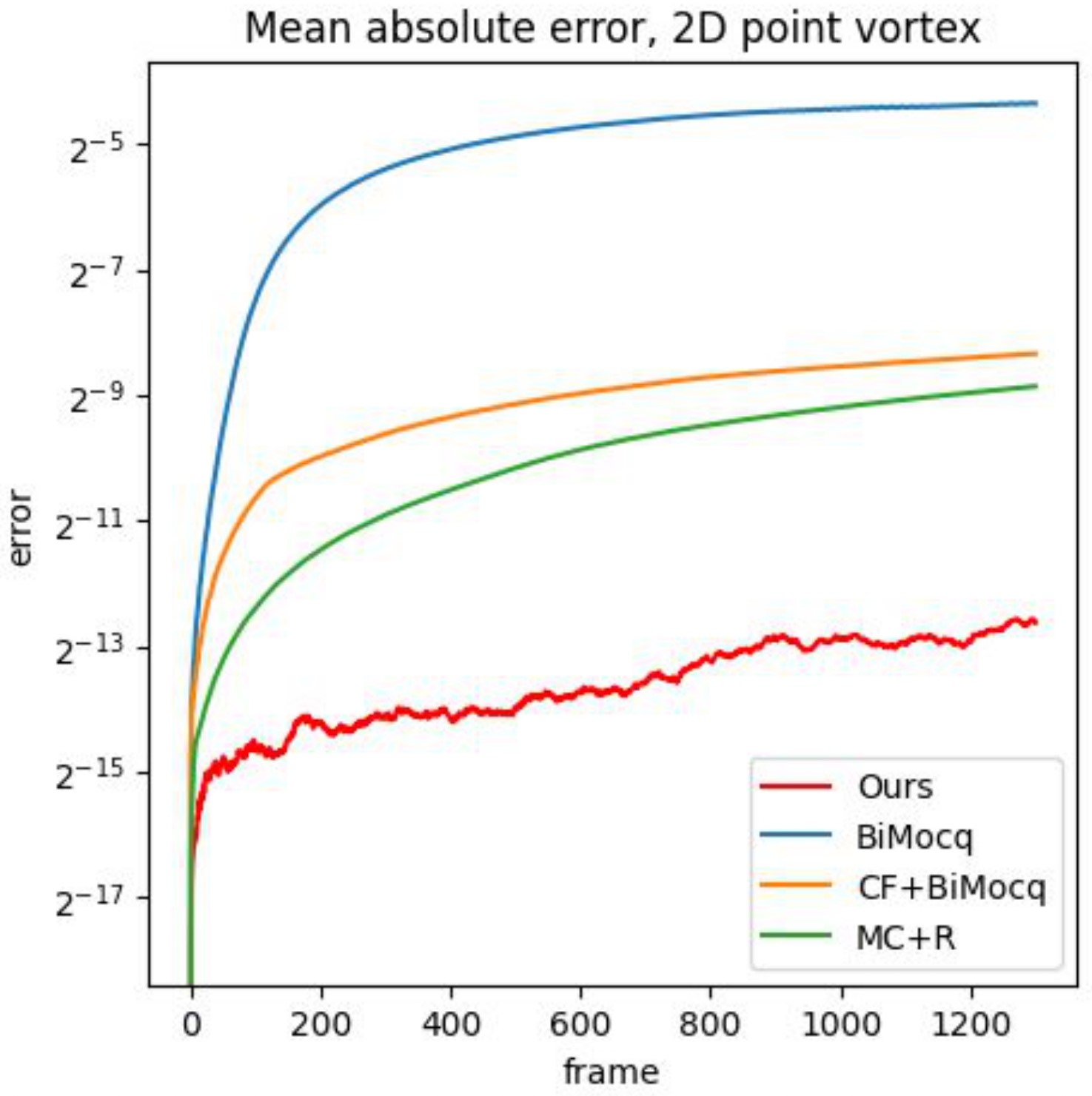}
 \caption{Left: the 3D kinetic energy plotted over time, which showcases the improved energy conservation property of \model. Right: mean absolute error of \model over time compared to the benchmarks in a steady 2D flow.}
 \label{fig:energy_and_error}
 \vspace{-18pt}
\end{figure}

\paragraph{2D Analysis: Steady Point Vortex}
An isolated point vortex in the absence of viscosity or external forces is a steady state where the velocity field remains constant over time. This is nevertheless challenging to satisfy for simulation methods due to numerical dissipation. 
Here, we compare \model to BiMocq \cite{qu2019efficient}, CF+BiMocq \cite{nabizadeh2022covector}, and MC+R \cite{zehnder2018advection}, three of the advanced simulation methods recently proposed, in terms of their adherence to the steady state after long simulations. The mean absolute error between the simulated velocity and the steady velocity over 1300 frames is plotted in Figure~\ref{fig:energy_and_error} (right), and the final error image is depicted in Figure~\ref{fig:two_2d_comparisons} (right). 
It can be seen that CF+BiMocq and MC+R both yield much reduced errors compared to BiMocq, with MC+R being slightly better than CF+BiMocq. Our method outperforms these two methods by an order of magnitude. 

It is particularly interesting to note that, with the three methods all built upon the premise of reduced interpolation errors with long-range mapping, BiMocq and CF+BiMocq yield less accurate results than the single-step MC+R method, while \model yields a much more accurate one than MC+R. This corroborates the argument that the potential of flow map-based advection can only be realized when the flow maps are simultaneously 1) long-range, and 2) accurate, which reiterates the necessity of our bidirectional marching scheme.

\begin{figure*}[t]
 \centering
 \includegraphics[width=.985\textwidth]{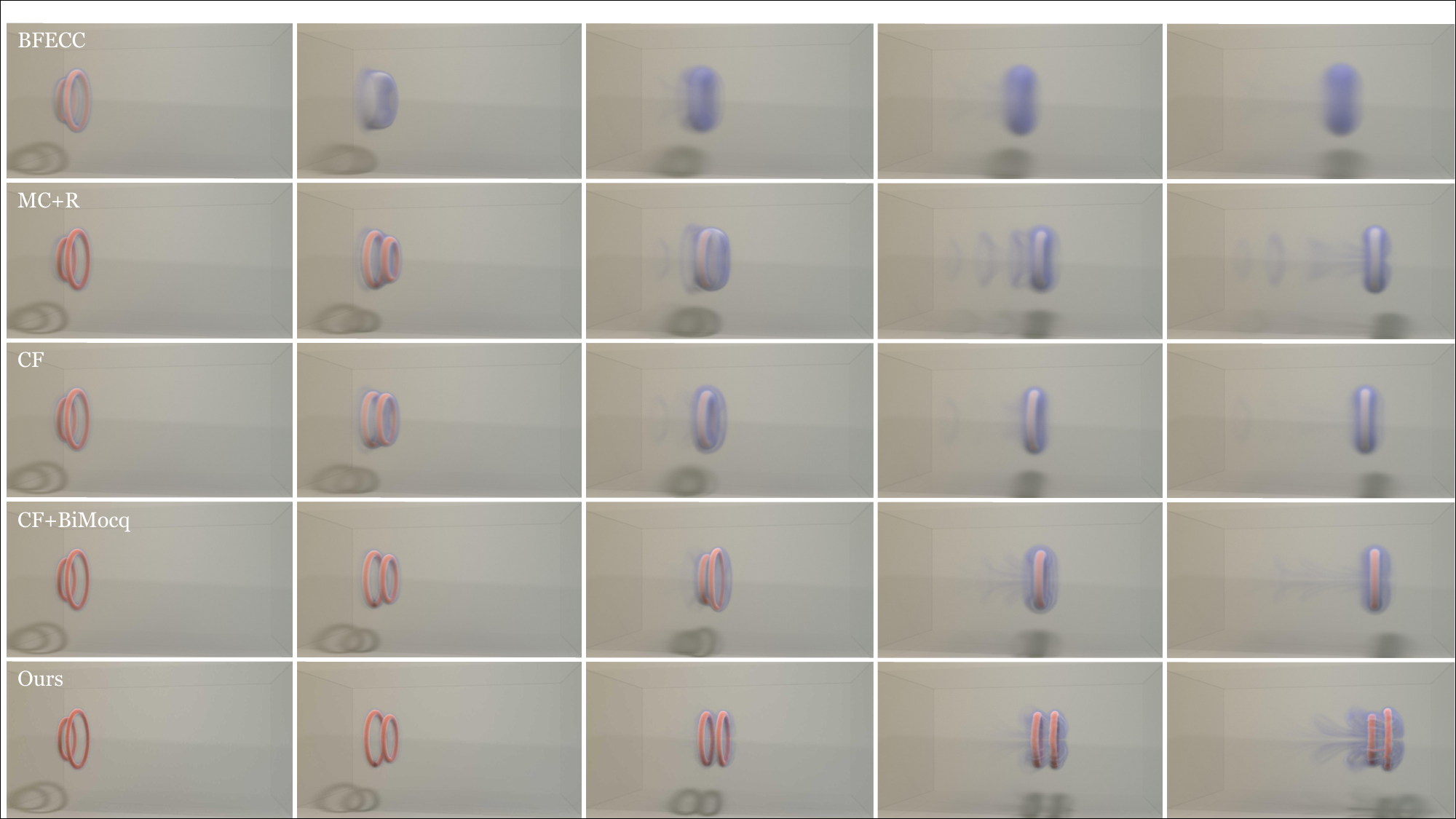}
 \caption{Comparison of 3D leapfrogging vortices to the benchmarks. Our method simulates vortex rings that remain separate even after the fifth leap, while those simulated by the benchmarks merge after at most the third leap, showing improved correspondence to the expected phenomenon.}
 \label{fig:leapfrog}
\end{figure*}

\paragraph{3D Analysis: Vortex Rings} As shown in Figure~\ref{fig:leapfrog}, we set up the leapfrogging vortex rings experiment and compare \model to \rev{4} benchmarks: BFECC \cite{kim2006advections}, \rev{MC+R \cite{narain2019second}}, CF and CF+BiMocq \cite{qu2019efficient, nabizadeh2022covector}. The time-varying kinetic energy for all five methods are plotted in Figure~\ref{fig:energy_and_error} (left), and it can be observed that our method conserved energy better than its counterparts. It is known that in the conservative case, the parallel vortex rings will remain separated and leap around each other indefinitely. As shown in Figure~\ref{fig:leapfrog}, our improved energy conservation indeed translates to visibly better fulfillment of this phenomenon, as \model yields vortex rings that remain separate after the fifth leap, while those from the other methods merge after at most three leaps. As shown in Figure~\ref{fig:two_2d_comparisons} (left), we also conduct an analogous experiment in 2D and achieve similar results.

\begin{figure}
 \centering
 \includegraphics[width=.46\textwidth]{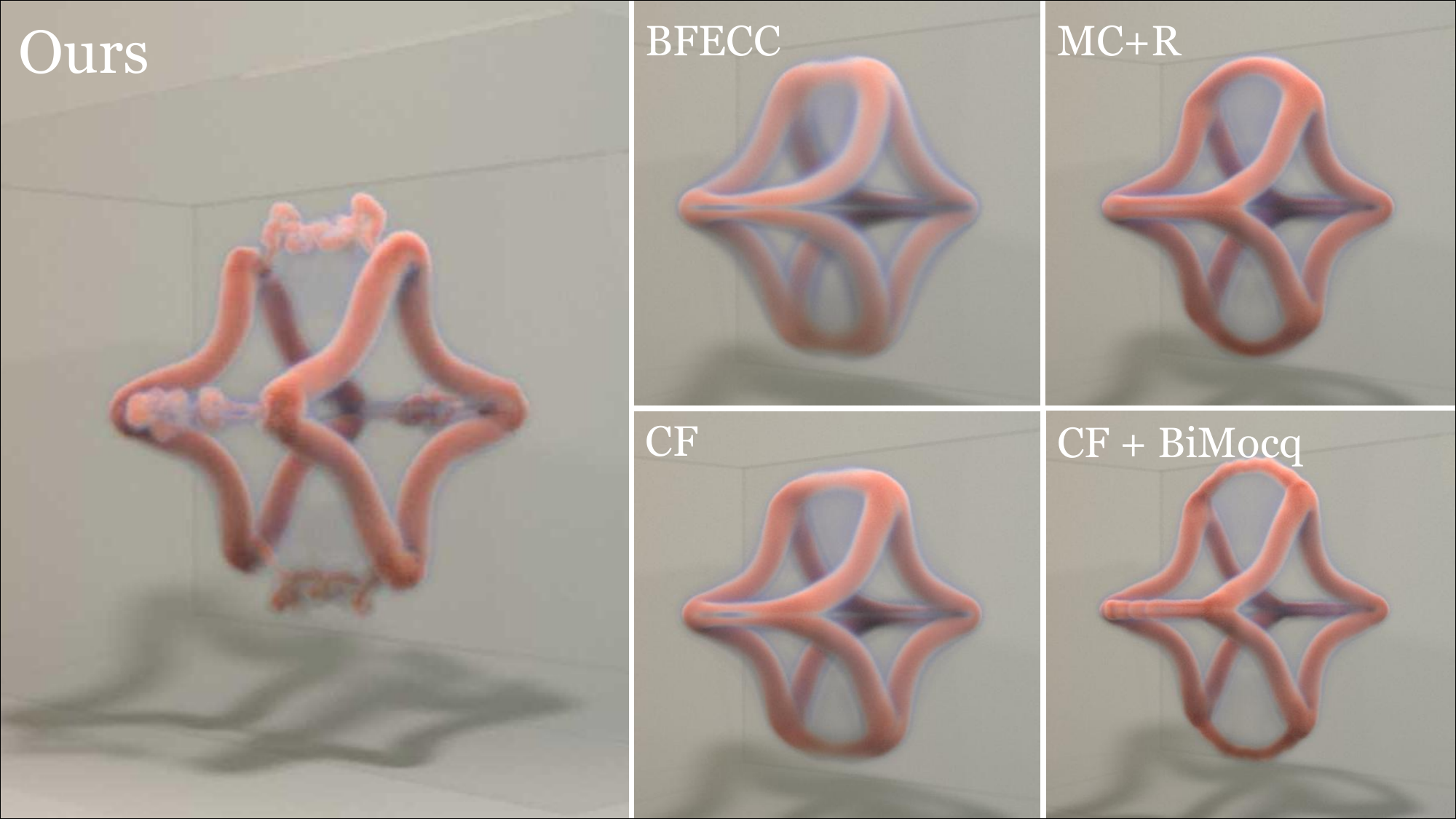}
 \caption{The vortex reconnecting instant of the ``four vortices'' example. Our method does not suffer from the numerical diffusion which manifests in the \textit{viscous bridges} between the two reconnected vortices.}
 \label{fig:four_corts_comp}
 \vspace{-3pt}
\end{figure}

Our method's reduction of numerical dissipation noticeably enhances the visual beauty of turbulent fluid simulation. For example, in Figure~\ref{fig:four_corts_comp}, we compare the vortex ring reconnection simulated by \model to that by the benchmarks. It can be observed that, the most diffusive method: BFECC creates damped, viscous bridges between the rings that keep them from reconnecting. The modern solvers (CF, \rev{MC+R} and CF+BiMocq) lessen the numerical diffusion to effectively thin these viscous bridges. \model offers another step up with its clean reconnection without the viscous bridges, which is more in line with the reference simulation \cite{Matsuzawa2022VideoTT}.

\setlength{\abovecaptionskip}{12pt}

\setlength{\abovecaptionskip}{12pt}

\setlength{\abovecaptionskip}{12pt}

\setlength{\abovecaptionskip}{8pt}

\paragraph{Influence of Reinitialization Steps}
\rev{
As shown in Figure~\ref{fig:hyperparameters_ablation} (left), we analyze the influence of the reinitialization step $n$ on the simulation accuracy by performing the aforementioned 2D steady-state test with $n=$ 1, 5, 9, 11, 13, 15, 17, 21, 25, 29, and 34. It can be seen that the error exhibits a u-shaped trend, and minimizes at $n = 17$. A small choice of $n$ can suffer from the frequent interpolations, while a large choice tends to suffer from the errors in long-range flow map marching. Currently, $n$ is empirically selected for each simulation.
}

\section{Examples}

With our high-performance \model simulator, we tackle a range of complex simulation scenarios, including vortex ring reconnections, vortex shedding from moving obstacles, and vortex development from fluid density difference. A catalog of our examples can be found in Table~\ref{tab: examples_table}. In our examples, we assume the shorter edge to have the unit length, and the setups are reported accordingly. \rev{We use a workstation with AMD Ryzen Threadripper 5990X and NVIDIA RTX A6000 to compute our examples.}

\paragraph{Leapfrogging Vortices (2D)} As shown in Figure~\ref{fig:two_2d_comparisons}, we conduct the classic 2D leapfrog experiment by placing four point vortices centered at $x$ = 0.25 and \rev{$y$ = 0.26, 0.38, 0.62, and 0.74}. The vortices have the same strength (magnitude) of 0.005, with the upper two being negative and the lower two being positive. The velocity fields are obtained from the point vortices using a mollified Biot-Savart kernel with support 0.02. The leapfrogging vortices will hit the right wall and separate into two vortex \textit{pairs} that will return to the left. Once these two pairs hit the left wall, they reassemble into the initial configuration and repeat the process. Our method is able to repeat this cycle 3 times with the vortices still separated.

\setlength{\abovecaptionskip}{12pt}
\begin{figure*}[t]
 \centering
 \includegraphics[width=.99\textwidth]{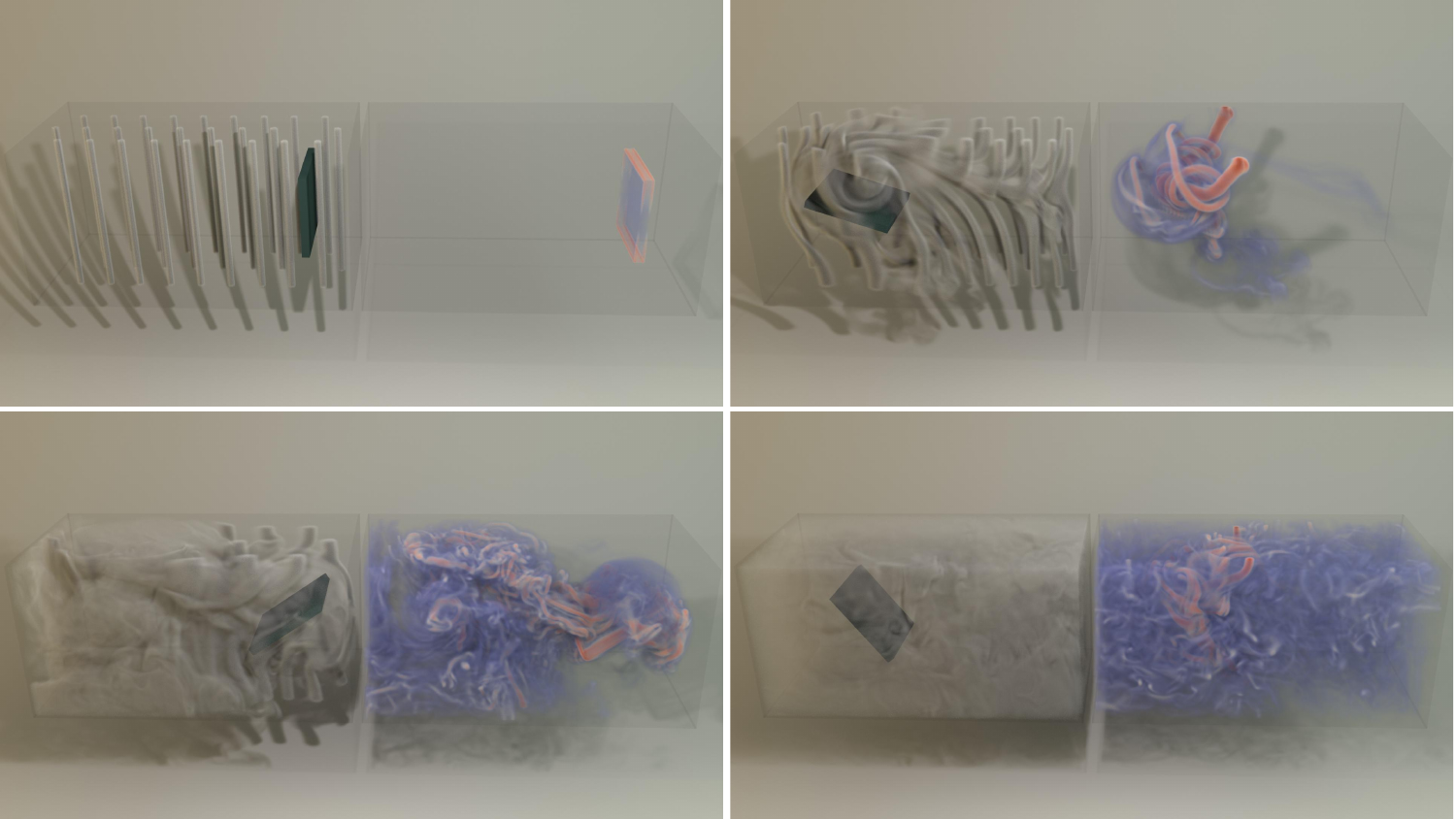}
 \caption{Turbulent flow induced by a rotating and translating paddle. The smoke is juxtaposed with the vorticity field that shows intricate vortex filaments.}
 \label{fig:moving_paddle}
\end{figure*}

\begin{figure*}
 \centering
 \includegraphics[width=.99\textwidth]{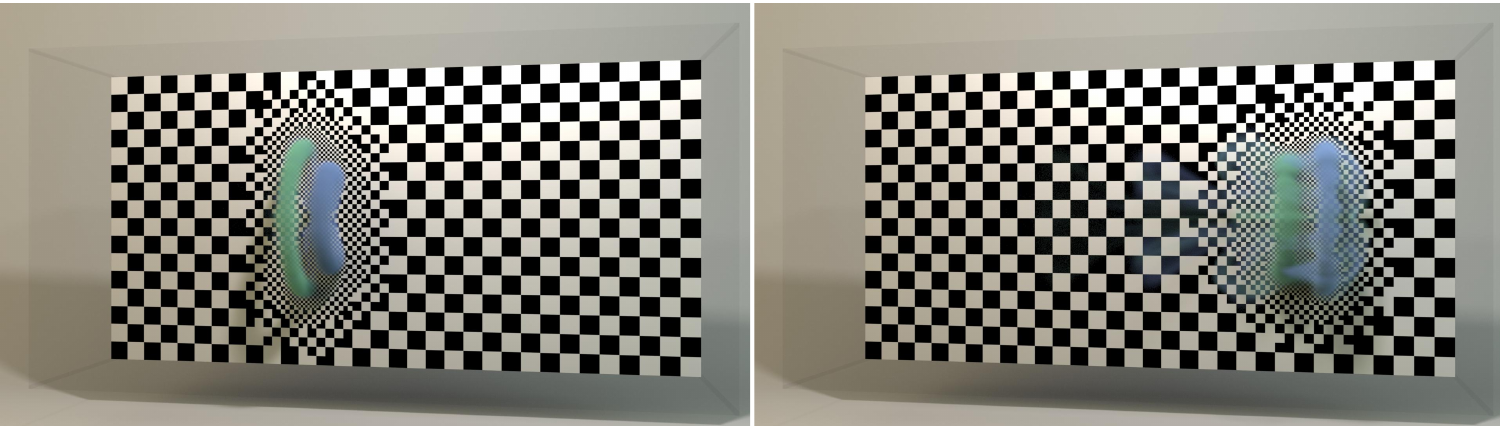}
 \caption{The evolving spatial discretization of our neural buffer $\mathcal{N}$'s spatially sparse feature grid, as the simulation proceeds.}
 \label{fig:sparse_grid_leapfrog}
\end{figure*}

\paragraph{Leapfrogging Vortices (3D)} As shown in Figure~\ref{fig:sparse_grid_leapfrog}, we initialize two parallel vortex rings with $x$ = 0.16 and 0.29125. The major radius is 0.21; the minor radius (the mollification support of the vortices) is 0.0168. Our method remains separate after the $5^\text{th}$ leap, while existing benchmarks diffuse and merge after at most the $3^\text{rd}$ one.

\paragraph{Oblique Vortex Collision} As shown in Figure~\ref{fig:oblique}, we initialize two vortex rings initially facing each other at the right angle. The center of the vortex rings are offset by 0.3 along the $x$-axis. The major radius of both is 0.13; the minor radius is 0.02. Upon collision, the vortices attach on the left side to form a single vortex ring, which gets catapulted to the right and divides into three smaller vortices.  

\paragraph{Headon Vortex Collision}
As shown in Figure~\ref{fig:headon},
we initialize two opposing vortex rings that are offset by 0.3 along the $x$-axis. The major radius is 0.065; the minor radius is 0.016. Upon collision, the two vortices stretch rapidly along the $yz$-plane while thinning along the $x$-axis, causing the structure to destabilize and break into a ring of small, secondary vortices facing radially outward, which resembles the experimental results by \citet{lim1992instability}.

\paragraph{Trefoil Knot}
As shown in Figure~\ref{fig:trefoil}, we reproduce with \model the classic trefoil knot setup by \citet{kleckner2013creation}. We simulate with the initialization file open-sourced by \citet{nabizadeh2022covector}, and observe that the knot structure correctly breaks into one larger vortex and one smaller vortex.

\paragraph{Four Vortices Collision} As shown in Figure~\ref{fig:four_vorts}, we initialize four colliding vortices following \citet{Matsuzawa2022VideoTT} by placing four vortex rings that each form right angles with its neighbors, essentially outlining a square in the $yz$-plane. The major radius is 0.13; the minor radius is 0.02.
The collision causes the four vortex rings to reconnect into two vortices shaped like four-pointed stars, leaving behind intricate, turbulent patterns. Both vortex rings then ``crawl'' towards the left and right walls, morphing their shapes. After hitting the walls, the two vortices separate into four vortex tubes.   

\paragraph{Moving Paddle} As shown in Figure~\ref{fig:moving_paddle}, we initialize arrays of smoke columns with zero initial velocity. A kinematic boundary rotates around the $z$-axis while translating along the $x$-axis. The translated position linearly interpolates between $x = 0.5$ and $x = 1.7$, with a time-dependent fraction $0.5\cdot(1+\cos(0.5\cdot t))$. The rotation angle relates to time as $0.75 \cdot t$. The paddle is a cube with a width and height of 0.54 and thickness of 0.05. The velocity difference between the paddle and the surrounding fluid creates vortex sheets that roll up into intricate vortex filaments. The vortical velocity field disintegrates the initial smoke columns into a turbulent mixture. 

\paragraph{Inkdrop} As shown in Figure~\ref{fig:teaser}, we initialize a vortex ring with a major radius of 0.06 and minor radius of 0.016, centered at $y$ = 1.85 facing the $-y$-direction. Without density difference, the vortex would simply translate downward. However, we give the smoke (which is similarly shaped as the vortex) a relative density of 1.7, and apply a gravitational force of $[0, -2, 0]$, which together elicits the Rayleigh-Taylor instability that deforms the vortex ring and evolves it into intricate vortex filaments.

\begin{figure*}[h]
    \begin{minipage}[t]{.72\textwidth}
    \captionsetup{type=table, position=above}
        \caption{\rev{The top compares our method's wall time to those of the benchmarks; the bottom breaks down the time cost into the different subroutines. Advect-\textit{T} and Advect-\textit{N} correspond to the \textit{traditional} and \textit{neural} aspects of the advection. The timings are obtained with a laptop with Intel Core i7-12700H and NVIDIA RTX 3070 Ti.}}
        \begin{tabularx}{\textwidth}{ Y | Y | Y |Y| Y|Y}
        \hlineB{3}
        \multicolumn{6}{c}{\textbf{\rev{Average Time Cost per Step}}}\\
        \hlineB{2.5}
        Method&BFECC&MC+R&CF&CF+BiMocq&Ours\\
        \hlineB{2.5}
        Time cost &0.509s&0.539s&0.525s&0.539s&9.01s\\
        \hlineB{2.5}
        \multicolumn{6}{c}{\textbf{\rev{Timing Breakdown of a Step}}}\\
        \hlineB{2.5}
        Project&Advect &Advect-\textit{T} & Advect-\textit{N} & Train & I/O \& etc.\\
        \hlineB{2.5}
        0.312s& 3.67s & 0.381s & 3.29s & 4.83s & 0.196s\\
        \hlineB{3}
        \end{tabularx}
        \label{table: timing}
        \bigskip
    \label{tab: speed_analysis}
    
    \end{minipage}
    \hfill
    \begin{minipage}[t]{.27\textwidth}
    \raisebox{-\height+0.4\baselineskip}{\includegraphics[width=\linewidth]{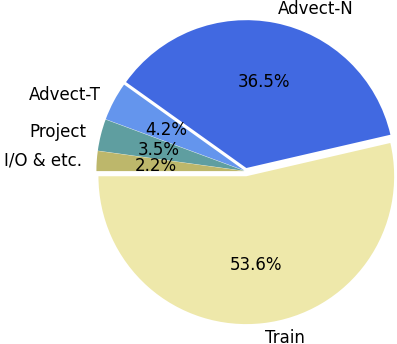}}
        \caption{Time cost breakdown.}
        \label{fig:timing_breakdown}
    \end{minipage}
    \vspace{0pt}
\end{figure*}

\section{Discussion and Future Works}
In this work, we propose \modelfull (\model), an effective simulation method based on our novel \inr neural representation, which bridges the mathematical models of characteristic mapping and impulse fluid mechanics with the efficacy of neural networks. 
Leveraging \inr's cutting-edge accuracy, speed, and memory efficiency, we compute highly accurate bidirectional flow maps at a viable memory cost, to facilitate \model's exceptional computational capabilities as demonstrated in a variety of challenging simulation scenarios, showcasing state-of-the-art energy conservation, visual complexity, and accurate recreation of real-world phenomena. 

Our work represents a significant step towards harnessing the power of machine learning for high-fidelity, first-principled simulation. 
Leveraging the virtues of neural networks, our work presents a general approach for efficiently storing high-dimensional fields, which unlocks a myriad of algorithmic designs that were previously unattainable, and highlights the immense potential of using INRs as data primitives in physics-based simulation. On a higher level, our method differentiates from prior {neural simulation} methods that focus on the emulation of non-neural methods, by leveraging neural techniques to \textit{extend the frontier} attained by existing schemes into new, unknown territories. It thereby offers a new perspective on the incorporation of machine learning in numerical simulation research for computer graphics and computational sciences alike.

\begin{table*}[t]
\centering\small
\begin{tabularx}{\textwidth}{Y | Y | Y | Y | Y | Y | Y | Y | Y |Y| Y}
\hlineB{3}
Name & Figure & Resolution & CFL & Reinit. Steps \rev{$n$} & Encoder Min. Res & Encoder Max Res & Max Num. Params. & Avg. Num. Params. & Max Training Iters & Batch Size\\
\hlineB{2.5}
2D Leapfrog & Figure~\ref{fig:two_2d_comparisons} & 256 $\times$ 256 & 1.0 & 20 & 32 $\times$ 32 & 256 $\times$ 256 & 108,272 & 75,848 & 3,000 & 25,000 \\
\hlineB{2}
3D Leapfrog & Figure~\ref{fig:sparse_grid_leapfrog} & 256 $\times$ 128$\times$ 128 & 0.5 & 20 & 32 $\times$ 16 $\times$ 16 & 256 $\times$ 128$\times$ 128 & 4,721,104 & 2,905,837 & 3,000 & 240,000 \\
\hlineB{2}
3D Oblique & Figure~\ref{fig:oblique}  &128 $\times$ 128$\times$ 128 & 0.5 & 20 & 16 $\times$ 16 $\times$ 16 & 128 $\times$ 128$\times$ 128 & 1,828,192 & 1,359,180 & 3,000 & 80,000 \\
\hlineB{2}
3D Headon & Figure~\ref{fig:headon} & 128 $\times$ 256$\times$ 256 & 0.5 & 12 & 16 $\times$ 32 $\times$ 32 & 128 $\times$ 256$\times$ 256 & 16,623,504 & 10,387,849 & 3,000 & 240,000 \\
\hlineB{2}
3D Trefoil & Figure~\ref{fig:trefoil}  & 128 $\times$ 128$\times$ 128 & 0.5 & 20 & 16 $\times$ 16 $\times$ 16 & 128 $\times$ 128$\times$ 128 & 2,587,280 & 1,693,634 & 3,000 & 80,000 \\
\hlineB{2}
3D Four Vortices & Figure~\ref{fig:four_vorts} & 256 $\times$ 128$\times$ 128 & 0.5 & 20 & 32 $\times$ 16 $\times$ 16 & 256 $\times$ 128$\times$ 128 & 4,306,160 & 1,725,884 & 1,500 & 240,000 \\
\hlineB{2}
3D Paddle & Figure~\ref{fig:moving_paddle} & 256 $\times$ 128$\times$ 128 & 0.5 & 8 & 32 $\times$ 16 $\times$ 16 & 256 $\times$ 128$\times$ 128 & 6,036,240 & 2,714,730 & 500 & 240,000 \\
\hlineB{2}
3D Inkdrop & Figure~\ref{fig:teaser} & 256 $\times$ 512$\times$ 256 & 0.5 & 12 & 64 $\times$ 32 $\times$ 32 & 256 $\times$ 512$\times$ 256 & 30,044,000 & 18,193,506 & 1,000 & 320,000 \\
\hlineB{3}
\end{tabularx}
\vspace{5pt}
\caption{The catalog of all our 2D and 3D simulation examples. \rev{The exact numbers of simulation steps and training iterations are determined dynamically based on early-termination and the CFL condition. Typically, each example consists of 300-500 frames, a frame requires $5-10$ steps, and each step requires $100-500$ training iterations.}}
\label{tab: examples_table}
\end{table*}

Our work is subject to several limitations. First, despite \inr's state-of-the-art efficiency, our neural advection scheme still incurs a major performance bottleneck compared to traditional methods. 
\rev{
As shown in Table~\ref{tab: speed_analysis} and Figure~\ref{fig:timing_breakdown}, compared to BFECC, MC+R, and CF, our neural method increases the overall wall time by one order of magnitude, with over $90\%$ spent on neural-related operations. To better handle larger-scale simulations, future research might further reduce the INR's training time, or devise time integration schemes that train more sporadically.
}
Furthermore, our algorithm \rev{currently only treats smoke simulation. To extend it for water simulation, additional buffering techniques and advanced interface representations are called for to handle viscosity and surface tension respectively}. Beyond fluid systems, our flow map method's potential in simulating solids and multi-physics systems might also be investigated to open up an even broader range of applications.

\begin{acks}
We thank the anonymous reviewers for their insightful feedback.
We thank Shiying Xiong, Yuchen Sun, and Duowen Chen for the valuable discussions.
Georgia Tech and Dartmouth authors acknowledge NSF IIS \#2313075, ECCS \#2318814, CAREER \#2144806, IIS \#2106733, OISE \#2153560, and CNS \#1919647 for funding support. Stanford authors are supported in part by NSF RI \#2211258 and Google. We credit the Houdini education license for producing the video animations.
\end{acks}

\bibliographystyle{ACM-Reference-Format}
\bibliography{refs_ML_sim.bib, refs_INR.bib, refs_flow_map.bib, refs_simulation.bib}

\clearpage
\appendix
\section{Additional Pseudocode}
\label{sec:additional_pseudocode}
In this section, we provide additional pseudocodes to supplement Section~\ref{sec:simulation}.

Algorithm~\ref{alg:RK4} details the procedure for our custom RK4 integration scheme for evolving the flow maps and flow map Jacobians, Algorithm~\ref{alg:midpoint} outlines the second-order, midpoint method, and Algorithm~\ref{alg:bfecc} describes the error-compensated impulse advection scheme. 

\begin{algorithm}
\caption{Interleaved RK4 for $\phi$ and $\mathcal{F}$}
\label{alg:RK4}
\begin{flushleft}
        \textbf{Input:} $\bm{u}$, $\phi$, $\mathcal{F}$, $\Delta t$~\\
        \textbf{Output:} $\phi_\text{next}$, $\mathcal{F}_\text{next}$
\end{flushleft}
\begin{algorithmic}[1]
\State $(\bm{u}_1, \nabla \bm{u}\vert_1) \gets \textbf{Interpolate}(\bm{u}, \phi)$;
\State $\frac{\partial \mathcal{F}}{\partial t}\vert_1 \gets \nabla \bm{u}\vert_1 \mathcal{F}$;
\State $\phi_1 \gets \phi + 0.5 \cdot \Delta t \cdot \bm{u}_1$;
\State $\mathcal{F}_1 \gets \mathcal{F} + 0.5 \cdot \Delta t \cdot \frac{\partial \mathcal{F}}{\partial t}\vert_1$;
\State $(\bm{u}_2, \nabla \bm{u}\vert_2)\gets \textbf{Interpolate}(\bm{u}, \phi_1)$;
\State $\frac{\partial \mathcal{F}}{\partial t}\vert_2 \gets \nabla \bm{u}\vert_2 \mathcal{F}_1$;
\State $\phi_2 \gets \phi + 0.5 \cdot \Delta t \cdot \bm{u}_2$;
\State $\mathcal{F}_2 \gets \mathcal{F} + 0.5 \cdot \Delta t \cdot \frac{\partial \mathcal{F}}{\partial t}\vert_2$;
\State $(\bm{u}_3, \nabla \bm{u}\vert_3)\gets \textbf{Interpolate}(\bm{u}, \phi_2)$;
\State $\frac{\partial \mathcal{F}}{\partial t}\vert_3 \gets \nabla \bm{u}\vert_3 \mathcal{F}_2$;
\State $\phi_3 \gets \phi + \Delta t \cdot \bm{u}_3$;
\State $\mathcal{F}_3 \gets \mathcal{F} + \Delta t \cdot \frac{\partial \mathcal{F}}{\partial t}\vert_3$;
\State $(\bm{u}_4, \nabla \bm{u}\vert_4)\gets \textbf{Interpolate}(\bm{u}, \phi_3)$;
\State $\frac{\partial \mathcal{F}}{\partial t}\vert_4 \gets \nabla \bm{u}\vert_4 \mathcal{F}_3$;
\State $\phi_\text{next} \gets \phi + \Delta t \cdot \frac{1}{6}\cdot (\bm{u}_1 + 2 \cdot \bm{u}_2 + 2 \cdot \bm{u}_3 + \bm{u}_4)$;
\State $\mathcal{F}_\text{next} \gets \mathcal{F} + \Delta t \cdot  \frac{1}{6} \cdot (\frac{\partial \mathcal{F}}{\partial t}\vert_1 
 + 2 \cdot \frac{\partial \mathcal{F}}{\partial t}\vert_2 + 2 \cdot\frac{\partial \mathcal{F}}{\partial t}\vert_3 + \frac{\partial \mathcal{F}}{\partial t}\vert_4)$;
\end{algorithmic}
\end{algorithm}

\begin{algorithm}
\caption{Midpoint Method}
\label{alg:midpoint}
\begin{flushleft}
        \textbf{Input:} $\bm{u}$~\\
        \textbf{Output:} $\bm{u}_\text{mid}$
\end{flushleft}
\begin{algorithmic}[1]
\State Reset $\psi, \mathcal{T}$ to identity;
\State March $\psi, \mathcal{T}$ with $\bm{u}$ and $-0.5\Delta t$ using Alg.~\ref{alg:RK4};
\State $\bm{m}_\text{mid} \gets \mathcal{T}^T\bm{u}(\psi)$;
\If{use external force}
\State ${\bm{m}}_\text{mid} \gets {\bm{m}}_\text{mid} + 0.5\cdot\Delta t \cdot \bm{f}_\text{ext};$
\EndIf
\State $\bm{u}_\text{mid} \gets \textbf{Poisson}({\bm{m}}_\text{mid} )$;
\end{algorithmic}
\end{algorithm}

\begin{algorithm}
\caption{Error-compensated Impulse Advection}
\label{alg:bfecc}
\begin{flushleft}
        \textbf{Input:} $\bm{u}_0$, $\psi$, $\mathcal{T}$, $\phi$, $\mathcal{F}$~\\
        \textbf{Output:} $\bm{u}$
\end{flushleft}
\begin{algorithmic}[1]
\State $\bar{\bm{m}} \gets \mathcal{T}^T\bm{u}_0(\psi)$;
\State $\bar{\bm{m}}_0 \gets \mathcal{F}^T\bar{\bm{m}}(\phi)$;
\State $\bm{e} \gets 0.5 \cdot (\bar{\bm{m}}_0 - \bm{u}_0)$;
\State $\bar{\bm{e}} \gets \mathcal{T}^T\bm{e}(\psi)$;
\State $\hat{\bm{m}} \gets \bm{\bar{m}} - \bar{\bm{e}}$;
\State $\bm{m} \gets \textbf{Clamp}(\hat{\bm{m}})$;
\State $\bm{u} \gets \textbf{Poisson}(\bm{m})$;
\end{algorithmic}
\end{algorithm}

\section{Implementation Details for \model}
\label{sec:sim_implementation}

\paragraph{MAC Grid Stencil}
We use the standard MAC grid \cite{harlow1965numerical} for storing the velocity components, and we evolve flow maps and flow map Jacobians on face centers. To carry out impulse-based advection, the matrix multiplication $\mathcal{T}^{T}\bm{u}$ needs to be evaluated, which can be rewritten as \rev{$[\frac{\partial \psi}{\partial x} \cdot \bm{u}, \frac{\partial \psi}{\partial y} \cdot{\bm{u}}, \frac{\partial \psi}{\partial z} \cdot{\bm{u}}]^T$}. As a result, for faces storing $\bm{u}_x$, instead of storing the full Jacobian matrix $\mathcal{T}$, we only need to store its first column $\frac{\partial \psi}{\partial x}$. Similarly, for faces with $\bm{u}_y$, we store $\frac{\partial \psi}{\partial y}$; for faces with $\bm{u}_z$, we store $\frac{\partial \psi}{\partial z}$.
The situation is analogous for $\phi$ and $\mathcal{F}$ as well. 

As a result, in our implementation, the $\mathcal{T}$ and $\mathcal{F}$ matrices are always stored as columns at the staggered face centers. And the Algorithm~\ref{alg:RK4} is applied to evolve each column individually.

\paragraph{Interpolating $\bm{u}$ and $\nabla \bm{u}$}
We approximate the velocity $\bm u$ and velocity Jacobian $\nabla \bm{u}$ at subgrid points using the MPM interpolation scheme \cite{jiang2016material} with the quadratic kernel:
\begin{equation}
N(x)=
\begin{cases}
    \frac{3}{4} - \vert x \vert^2 & 0 \leq \vert x \vert < \frac{1}{2},\\
    \frac{1}{2}(\frac{3}{2}-\vert x \vert)^2 & \frac{1}{2} \leq \vert x \vert < \frac{3}{2},\\
    0    & \frac{3}{2} \leq \vert x \vert.
\end{cases}
\end{equation}
Empirically, we find the quadratic kernel to perform the best, as the linear kernel leads to numerical instabilities due to its discontinuous gradient and the cubic kernel leads to numerical diffusion due to its larger support size. Alternatively, automatic differentiation can be used to directly query velocity gradients from the \inr buffer $\mathcal{N}$. In practice, however, this leads to significant noise in the gradient computation, and makes the simulation unstable.  

\paragraph{Sizing Function}
We compute the sizing value $S$ as the Frobenius norm of the velocity Jacobian:
\begin{equation}
\label{eq:sizing}
S = \sqrt{\sum_{i=1}^d\sum_{j=1}^d \bigg\vert\frac{\partial\bm{u}_i}{\partial \bm{e}_j}\bigg\vert^2}.
\end{equation}
Activating the feature grid of $\mathcal{N}$ using $S$ computed in this way can lead to sharp level transitions. Our computational scheme handles these sharp transitions naturally, but they nevertheless undermine the fitting accuracy. To encourage a smooth transition between levels, we dilate the sizing function for 1024 iterations as follows:
\begin{equation}
    S_{i,j}^{k+1} =\max(S_{i,j}^{k}, 0.25 \cdot (S_{i-1,j}^{k} + S_{i+1,j}^{k} + S_{i,j-1}^{k} + S_{i,j+1}^{k})).
\end{equation}
In other words, we allow the sizing value in a voxel to diffuse to neighboring voxels without decreasing its original value.

\paragraph{Gravity and Buoyancy} For the realization of density difference-driven effects, we apply a Boussinesq buoyancy force $\bm{f}_\text{ext} = c \cdot \rho \cdot \bm{g}$. We assume $\rho$ remains constant during flow paths, so the integration for this term can be trivially computed by multiplying with the total timelapse. We note that such a force integral should be added to the velocity $\bm{u}$ instead of the impulse $\bm{m}$. The raw velocity after this addition will be projected again to obtain the final velocity.

\paragraph{Solid Boundaries and Obstacles} Our \model simulation system naturally supports moving, voxelized boundaries. Since our neural buffer $\mathcal{N}$ stores spatiotemporal velocity fields that respect the moving boundary's geometry and kinematics, a path obtained by marching $\mathcal{N}$ will be valid and will not penetrate the boundary as long as $\mathcal{N}$ has stored the velocity accurately, which ensures that our advection scheme will work in this case just like in the boundary-free cases. 

With this being said, it is indeed more challenging for $\mathcal{N}$ to learn the velocity field when obstacles are present, as they lead to sharp transitions in the velocity profile near the edges. We opt to smooth out these sharp velocity transitions by extrapolating the fluid velocity into the boundary, and letting $\mathcal{N}$ learn the extrapolated version, which significantly improves the convergence performance. We note that for the purpose of flow map-marching, modifying the {solid} velocity does not sacrifice the physical integrity, because a \textit{valid} path makes no use of the solid velocity anyways, so it would not be affected by such a modification; a path that penetrates into the boundary would be \textit{invalid} to begin with, and in that case, it can only benefit from such a modification as the marched flow map would be more smooth, which makes the simulation more stable. For Poisson solving, we make no modifications to the velocity.     

Additionally, with a moving, voxelized boundary inside all Neumann walls, the total numerical divergence can deviate from 0, causing the Poisson solver to not converge. We compensate for this deviation by offsetting the divergence of each fluid voxel by a constant value so that the divergence field sums to 0.

\section{Setup Details for Subsection~\ref{subsec:inr_validation}}
\label{sec:experimental_setup}
\paragraph{Experimental Setup}
We set up the benchmarking experiment by generating offline a 25-frame sequence of MAC grid velocity fields for both 2D and 3D. In 2D, we use the leapfrog setup; in 3D, we use the trefoil knot setup. For all four methods, we adjust the hyperparameters so that the parameter count is about the same. For fairness of comparison, all three benchmarks are extended with the multi-branch decoder in our method, one for each spatial dimension. All methods are trained for 2000 iterations with the same optimizer and learning rate scheduler. In 2D, all methods are trained with a batch size of 25000; in 3D, all methods are trained with a batch size of 120000.

\paragraph{INGP} For the 2D test, INGP uses 16 scales from min resolution [16, 16, 2] to max resolution [256, 256, 24] (the last dimension is time). The max number of parameters for each level is set to 2600. For the 3D test, INGP uses 16 scales from min resolution [8, 8, 8, 2] to max resolution [256, 256, 256, 24] with the max number of parameters being 87880 per level. In both 2D and 3D, we use an ensemble of MLP decoders
with a depth of $2$ and a width of $64$.

\paragraph{KPlanes} For the 2D test, KPlanes uses 3 scales from min resolution [16, 16, 8] to max resolution [64, 64, 8] (the last dimension is time), using no multi-resolution for the time axis as suggested \rev{by \citet{fridovich2023k}}. The feature length is set to 12, which is smaller than the paper's suggested number to stay within a comparable budget. For the 3D test, KPlanes uses 3 scales from [32, 32, 32, 8] to [128, 128, 128, 8], and a feature length of 32. The decoders used for KPlanes are the same as those for INGP.

\paragraph{SIREN} For the 2D test, SIREN uses an MLP with a depth of $4$ and a width of $180$. For the 3D test, SIREN uses an MLP with a depth of $8$ and a width of $256$. Both MLPs have sinusoidal activation functions. Following \rev{the approach by \citet{sitzmann2020implicit}}, we employ a frequency multiplier of $\omega_0 = 30$ to boost up the networks' characteristic frequency. We perform this for the spatial axes only to ensure temporal smoothness. 

In order to fairly compare the computation time, we implement all four methods using PyTorch for neural networks and optimization, and Taichi \cite{hu2019taichi} for efficient feature vector storage.

The training procedure in this test mimics the actual simulation scenario, where our ``dynamic scene'', a 25-frame velocity sequence, is not presented to the INRs all at once, but rather as a stream of frames. Each model is only presented with the newest frame, and the prior frames must be read from its auxiliary buffer $\hat{\mathcal{N}}$.  
\end{document}